\documentstyle[preprint,aps,epsf,floats,rotating]{revtex}

\tightenlines
\newcommand{\beq}{\begin{equation}}
\newcommand{\eeq}{\end{equation}}
\newcommand{\bqa}{\begin{eqnarray}}
\newcommand{\eqa}{\end{eqnarray}}
\newcommand{\nn}{\nonumber}

\def\square{\vcenter{\vbox{\hrule height.4pt
          \hbox{\vrule width.4pt height8pt
          \kern8pt\vrule width.4pt}\hrule height.4pt}}}

\begin{document}
\preprint{
\vbox{\halign{&##\hfil\cr
        & hep-th/0401171 \cr
&\today\cr }}}

\title{Short-range Interaction and  Nonrelativistic $\phi^4$ Theory in Various Dimensions}

\author{Yu Jia}
\address{Department of Physics and Astronomy, 
Michigan State University, East Lansing MI 48823}

\maketitle

\begin{abstract}
{\footnotesize 
We employ the effective field theory method
to systematically study the  short-range interaction  in two-body sector
in 2, 3 and 4 spacetime dimensions, respectively.
The $\phi^4$ theory is taken as a specific example and
matched onto the nonrelativistic effective theory to one loop level.
An exact, Lorentz-invariant expression for the S-wave amplitude is presented,
from which  the nonperturbative information can be easily extracted.
We pay particular attention to the renormalization group analysis in the 3 dimensions,
and show that  relativistic effects qualitatively change the renormalization group
flow of  higher-dimensional operators.
There is one ancient claim that triviality of the 4-dimensional 
$\phi^4$  theory can be substantiated in the nonrelativistic limit.  
We illustrate that this assertion arises from  treating  the 
interaction between two nonrelativistic particles as literally zero-range,
which is incompatible  with the Uncertainty Principle.
The S-wave effective range in this theory  is identified
to be approximately $16/3\pi$ times the Compton wavelength.
}\end{abstract}

\newpage

\section{Introduction}
\label{intro}

Short-range forces commonly arise in nuclear and condensed-matter system.
Because the microscopic dynamics is often not well understood, 
one usually models the potential with some arbitrary parameters
based on  empirical assumptions.
These parameters are then adjusted  by  trial and error 
from numerical solution of the Schr\"{o}dinger equation.

Because of the  {\it ad hoc} nature in  constructing the potential, 
this traditional  method suffers some severe shortcomings. 
Most notably,  it involves the uncontrolled approximations,  
thus making a reasonable error estimate impossible.

These difficulties can be overcome if the effective field theory (EFT) approach is instead exploited.
EFT provides a model-independent means to address  problems with separated scales 
(For a modern and comprehensive review on this topic, see Ref.~\cite{Georgi:qn}).
When nonrelativistic particles interact through short-range forces,
their de Broglie wavelengths are  much longer than the typical range of interaction.
Therefore, they can be treated as point particles,
and the low energy dynamics can be described by  a local nonrelativistic EFT.
The coexistence of two disparate scales,  the momentum $k$ and the cutoff $\Lambda$,  
which is roughly the inverse of the interaction range, 
validates a systematic expansion in  powers of $k/\Lambda$. 
Contrary to phenomenological potential model, this framework 
accommodates a transparent power counting, 
so that the error estimate can be performed systematically.
Recently, this method has been fruitfully applied to the few nucleon system~\cite{Beane:2000fx}
and cold dilute Bose gas~\cite{Braaten:1996rq} and Fermion gas~\cite{Hammer:2000xg}.

The $\delta$-function potential is usually considered as the prototype of
the short-range interaction~\cite{Jackiw:BEG,Manuel:1993it,Adhikari:1995uu}.
Simple enough as it may look,
this zero-range potential turns out to be too singular when iterated in
the  Schr\"{o}dinger equation (except in one spatial dimension), 
so that one has to regulate and renormalize
the resulting ultraviolet divergences.

There is no systematic way to deal with renormalization in nonrelativistic quantum mechanics.
In contrast, EFT is the paradigm for the modern understanding of renormalization~\cite{Georgi:qn}.
Furthermore, 
the contributions suppressed by inverse powers of $\Lambda$,
and relativistic effects proportional to the powers of $k/m$, 
can be conveniently incorporated in this field-theoretic framework.
Neither of these can be easily coped with in the Schr\"{o}dinger formalism with contact potentials.

Short-range interaction in four dimensions  has  already been 
extensively explored in literature,
but not much effort has been devoted to the lower dimensional cases.
The goal of this paper is to employ the EFT method to systematically investigate 
the short-range interactions in the two-body sector in various spacetime dimensions.
The highlight of this work is to present an exact, Lorentz-invariant
expression for the S-wave scattering amplitude. 
This represents an improvement to the previous results, 
where the relativistic effects are usually ignored, or only incorporated to the first order.

The capability to arrive at an exact amplitude  
is rooted in the tremendous simplicity of  the nonrelativistic EFT, where one can
sum all the four-point Green functions analytically.
Some useful nonperturbative insight can be gained from this exact expression.
Specifically, this ability is indispensable when discussing the 2-body scattering in 2D and 3D,
where the amplitude at fixed order of perturbative expansion 
is  plagued by the zero-momentum singularity.
In contrast, taking into account  the singular terms to all orders,
this exact amplitude is well behaved in the $k \to 0$ limit.

It is always nice to have some concrete example at hand.
The  $\phi^4$ theory constitutes one simple, but instructive example.
In the  nonrelativistic limit, this theory  is expected to simulate
the  $\delta$-function potential. 
This consideration has motivated some discussions on 
different aspects of the $\phi^4$ theory 
in this limit~\cite{Beg:yh,Bergman:1991hf,Gomes:1996px,Consoli:1997ip}.
Unfortunately, each work is more or less isolated and the 
underlying rationale of EFT is not fully realized.

Unlike the nuclear or atomic system, where the microscopic dynamics is
too complicated to pinpoint analytically,
the full knowledge of this model theory allows us to determine 
all the EFT parameters to any desired order.
Benefiting from the power of EFT, we are able to develop some detailed understanding of
this theory in the nonrelativistic limit.
On the other hand, the $\phi^4$ theory also provides some useful guidances 
in deducing the Lorentz-invariant amplitude in the nonrelativistic EFT.

The benchmark feature of the  4D $\phi^4$ theory, as well as other non asymptotically-free theories, 
is  {\it triviality}, in the sense that 
the renormalized coupling $\lambda$  has to vanish, 
if one insists that this theory be valid all the way down to the arbitrarily  
short distance (say, Planck length)~\cite{Callaway:ya}. 
It is worth emphasizing,  this symptom needs not to be regarded as a serious trouble, 
since this theory has to merge into a more fundamental theory at some point 
long before approaching the Planck scale. 
As long as treated as a low energy effective theory,  
the predictivity of this theory is not sacrificed.

Nearly two decades ago, Beg and Furlong claimed
that  triviality of the $\phi^4$ theory can be explicitly
corroborated in the nonrelativistic limit~\cite{Beg:yh}. They regarded this
as another piece of evidence for the triviality,  
complimentary to the more rigorous ``proof"
from formulating this theory nonperturbatively on the spacetime lattice.
If their claim were true, it would open an easier way to access this 
rather formal problem.

This paper is organized as follows:
in Sec.~\ref{EL:tree}, 
we describe the most general  nonrelativistic effective Lagrangian 
which is relevant to the two-body S-wave scattering.
After the pitfall of the real $\phi^4$ theory in the nonrelativistic limit
is  pointed out, we  match this theory onto the EFT at the tree level. 
We also sketch how to implement relativistic corrections in the nonrelativistic EFT.
In Sec.~\ref{trivial}, we show that, contrary to what Beg and Furlong claimed,
the triviality of the 4D $\phi^4$ theory can not be substantiated in the nonrelativistic limit.
The flaw of their argument is traced, and attributed to the 
incorrect  renormalization of the power-law divergences in the cutoff scheme, 
or equivalently,  treating the two-body interaction in this theory literally 
as contact.

Sec.~\ref{matching:oneloop} is the main body of this paper,
in which  a detailed analysis of short-range forces in different space-time dimensions: 2, 3 and 4
is presented.  An exact, nonperturbative and Lorentz-invariant expression for the S-wave amplitude 
is obtained  when the dimensional regularization  is employed.
The $\phi^4$ theory is taken as a concrete example to illustrate the systematics of 
matching beyond tree level. 
The peculiarity for each spacetime dimensions is discussed.
In particular, we present a detailed renormalization group analysis for the 3D case, and
elaborate  on the role played by relativity.
We also identify the effective range in the 4D $\phi^4$ theory 
roughly to be the Compton wavelength.
We summarize our results in Sec.~\ref{summary}.

\section{Effective Lagrangian and Tree-Level Matching}
\label{EL:tree}

Before  moving on to the nonrelativistic EFT,
we first discuss some features of the  real $\phi^4$ theory.
It has the Lagrangian
\begin{eqnarray}
{\cal L} &=&  {1\over 2} \, \partial_\mu\phi \,\partial^\mu\phi -
{1 \over 2} \, m^2  \, \phi^2 - {\lambda \over 4!}\,\phi^4\,,    
\label{oldL}
\end{eqnarray}
where the natural unit $\hbar=c=1$ is adopted.
Vacuum stability requests a positive $\lambda$, 
corresponding to a repulsive contact interaction.
We don't specify the spacetime dimension 
in which (\ref{oldL}) is defined.

It should be cautioned,  the real $\phi^4$ theory 
is intrinsically  hostile to a nonrelativistic
description, 
because it doesn't  respect the particle number conservation.
A better candidate to study the nonrelativistic limit is
the complex $\phi^4$ theory  with a $U(1)$ charge~\cite{Beg:yh}.
Provided that one prepares $N$ nonrelativistic particles of the same charge,
excluding those of opposite charge, charge conservation
will guarantee that $N$ is conserved.

Lacking any continuous internal symmetry notwithstanding,
the real $\phi^4$ theory still possesses a legitimate
nonrelativistic limit for 2-body and 3-body interactions, 
thanks to the  energy-momentum conservation which forbids 
the number of particles to further decrease.

\begin{figure}[tb]
\centerline{\epsfysize=3.0truecm \epsfbox{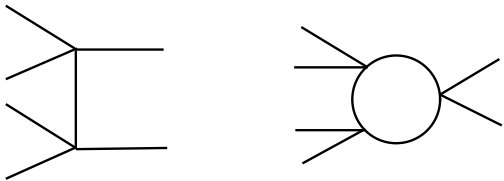}}
\noindent
\caption{ An example of particle number nonconservation 
in the real $\phi^4$  theory: four nonrelativistic particles can convert
into two relativistic ones.} 
\label{4to2} 
\vskip .2in
\end{figure}

However, it no longer makes sense to talk about the nonrelativistic
limit for more than 3 particles in this theory.
For example, Fig.~\ref{4to2} shows that even though all the four particles
in the initial state are nearly at rest,           
they can readily annihilate into two highly relativistic particles. 
There have been discussions about the many-body phenomena
of the real $\phi^4$  theory 
in the nonrelativistic limit~\cite{Consoli:1997ip}.
For the reason just outlined, the starting point of 
Ref.~\cite{Consoli:1997ip} looks rather questionable.

In this work,  we will confine ourselves in the 2-body nonrelativistic scattering,
so we are allowed to stay with the real $\phi^4$ theory.
For simplicity, we will focus on the S-wave (isotropic) scattering only, 
which is the most important partial wave in the low energy limit.

Symmetry is the guidepost in constructing a low energy effective theory.
For a nonrelativistic system, the most important symmetries are 
particle number conservation, Galilean invariance (${\bf k} \to {\bf k}+ m {\bf v}$),
time reversal and parity.
The most general form of the 
effective Lagrangian compatible with these symmetries is
\bqa
{\cal L_{\rm NR}} &=&  \Psi^* \, \left(i \partial_t + 
{\nabla^2 \over 2m} \right) \Psi
- {C_0 \over 4} (\Psi^* \Psi)^2
- {C_2 \over 8} \nabla(\Psi^* \Psi)\cdot \nabla(\Psi^* \Psi)
+\cdots\,.
\label{eq:effL}
\eqa
One can read off the nonrelativistic propagator to be 
$i/(E-{\bf k^2}/2m+i\epsilon)$, where $E$ is the kinetic energy,
and $\bf k$ is the 3-momentum.
In addition to the one-body kinetic operators,
we also include the two-body operators 
which describe the S-wave scattering.
Because Bose statistics and parity forbid
the  P-wave scattering (and all other odd partial waves),
the term proportional to $C_2$ is the only possible 
one with two gradients allowed by Galilean invariance.
The next tower of operators enter at 4th order of $\nabla$, 
which contribute to both the S- and D-wave scattering.

In an arbitrary reference frame, 
if two initial-state particles have momenta 
${\bf k}_1$  and ${\bf k}_2$, two final-state particles
carry momenta ${\bf k}_1^\prime$ and ${\bf k}_2^\prime$,
then the tree-level amplitude of (\ref{eq:effL}) reads
\bqa
{\cal A}_0^{\rm tree} & = & -C_0 -{C_2\over 4} [({\bf k}_1 -{\bf k}_1^\prime)^2 +
({\bf k}_1 -{\bf k}_2^\prime)^2 ]-\cdots \,,
\label{tree:amplitude}
\eqa
which is clearly Galilean-invariant.
In the center-of-momentum (C.M.) frame, the 3-momentum of each particle 
has the equal magnitude $k$, and the second term in the right-hand side
collapses to $-C_2 \, k^2$.
Thus we are reassured that the Lagrangian (\ref{eq:effL})
indeed describes the S-wave scattering.

The Wilson coefficients $C_0$, $C_2$, \ldots  encode all the short-distance
information. For two-nucleon system and Bose gas,  experimental input
is needed to deduce these coefficients.
For the $\phi^4$ theory, they can be determined from the {\it matching} procedure, 
{\it i.e.}, by requiring that
the effective theory (\ref{eq:effL}) reproduces the same physical observable 
as the full theory (\ref{oldL}),
up to a prescribed accuracy in powers of $k/\Lambda$,
order by order in loop expansion~\cite{Georgi:qn}.
Since the relativistic and nonrelativistic theories usually
adopt different conventions in  normalization of  states, 
one should be careful in specifying the matching condition.

To quickly access the nonrelativistic behavior of (\ref{oldL}),
it is customary to 
parameterize the relativistic field $\phi$ as~\cite{Beg:yh,Bergman:1991hf}
\bqa
\phi = {1\over \sqrt{2m}} \left( e^{-im t} \Psi +
e^{im t} \Psi^*\right) \,,
\label{reparameterize}
\eqa
where the field $\Psi$ only excites the nonrelativistic degree of freedom.
Plugging (\ref{reparameterize}) back into (\ref{oldL}), and dropping terms containing
the rapidly oscillating phases $\exp(\pm 2 i m t)$,
we obtain a new Lagrangian: 
\bqa
{\cal L}^\prime &=&  \Psi^* \, \left(i \partial_t + 
{\nabla^2 \over 2m} - {\partial_t^2 \over 2 m }\right) \Psi
- {\lambda \over 16m^2 } (\Psi^* \Psi)^2\,.
\label{new:matching:Lagr}
\eqa
The one-body operator with two time derivatives 
is not present in the standard nonrelativistic Lagrangian
(\ref{eq:effL}).
The function of this term is to recover the Lorentz symmetry.
To see this lucidly, one can rewrite the relativistic scalar propagator as
\bqa
{i\over k^2-m^2+i\epsilon} &=&  {1\over 2m}\,
{i\over E-{{\bf k}^2\over 2m}+{E^2\over 2m} +i\epsilon}\,,
\label{rewrite:propagator}
\eqa
where $E\equiv k^0-m$ is the kinetic energy.

The higher-derivative operators account for the dynamical short-distance effects, and 
their corresponding Wilson coefficients depend on the specific system under investigation.
In contrast, relativistic corrections represent a purely  kinematic effect,
thus  universal in any system.

In a realistic system,  relativistic effects are usually less important than
the higher-dimensional operators, because the cutoff $\Lambda$ is normally much less
than the particle mass.
For the $\phi^4$ theory,  we instead have $\Lambda \sim m$,
so the relativistic corrections, as well as
those terms proportional to powers of $k/\Lambda$, should be included simultaneously.

Relativistic effects can be systematically accounted for in the field-theoretic 
framework, which  sharply contrasts to the Schr\"{o}dinger formalism.
There are two different methods to implement relativistic corrections.  
One popular way is first redefining the field~\cite{Luke:1997ys}:
\bqa
 \Psi &=& 
 \left( 1+{\nabla^2 \over 4m^2}+{5 \nabla^4 \over 32 m^4} + \cdots \right) \Psi^\prime
 = \left( {m \over \sqrt{m^2+{\bf k}^2} } \right)^{1/2} \Psi^\prime \, ,
\label{field:redefine}
\eqa
and upon using  equation of motion,
one trades that extra operator with two time derivatives in (\ref{new:matching:Lagr}) 
for a infinite tower of terms with spatial gradients: 
\bqa
{\cal L}^{\prime\prime} &=&  \Psi^{\prime*} \, \left(i \partial_t + 
{\nabla^2 \over 2m} + {\nabla^4 \over 8 m^3 } + {\nabla^6 \over 16 m^5 }
+\cdots \right) \Psi^\prime
+\cdots\,.
\label{field:red:lag}
\eqa
From this equation, one can recover the relativistic dispersion relation
order by order:
\beq
 E \equiv  \sqrt{m^2+{\bf k}^2} -m = {{\bf k}^2\over 2m} - 
  {{\bf k}^4\over 8m^3} + { {\bf k}^6\over 16m^5} -\cdots \,.  
\label{rel:dispersion}
\eeq
In calculating relativistic corrections,  one replaces the kinetic part of
the effective Lagrangian (\ref{eq:effL}) by  (\ref{field:red:lag}).
These  higher-derivative one-body operators are understood to
be perturbatively inserted in the loop,  the energy-momentum relation 
for the external legs
should also be readjusted according to (\ref{rel:dispersion}).

However, as pointed out in Ref.~\cite{Chen:1999tn},
working with (\ref{field:red:lag}) is somewhat cumbersome,
because every term in the effective Lagrangian  (\ref{eq:effL})
is subject to the field redefinition (\ref{field:redefine}).
For example, the field redefinition exerting
on the lowest dimensional four-boson operator 
will induce  a new operator 
$-C_0/(8m^2)[(\nabla^2\Psi^{\prime*} \Psi^\prime)
\Psi^{\prime*}\Psi^\prime
+(\Psi^{\prime*} \nabla^2\Psi^\prime)\Psi^{\prime*}\Psi^\prime]$,
which  mixes with the  operator proportional to $C_2$. 
It  adds a new contribution $C_0/( 4m^2)({\bf k}_1^2 + {\bf k}_2^2
+{\bf k}_1^{\prime 2} +{\bf k}_2^{\prime 2})$ 
to the tree-level amplitude (\ref{tree:amplitude}).
In the C.M. frame,  the new amplitude becomes
\beq
{\cal A}_0^{\prime \, {\rm tree}}  = -C_0 \left(1- {k^2\over m^2} + \cdots \right) - C_2k^2 -\cdots.
\label{tree:ampl:2nd:rel}
\eeq

It turns out that directly replacing the kinetic part of (\ref{eq:effL}) 
by that of (\ref{new:matching:Lagr}),  without invoking the field redefinition (\ref{field:redefine}),
is much more convenient~\cite{Chen:1999tn}.
Firstly,  we are free from a plethora of induced new operators.
Secondly,  we just need insert that single one-body operator with two time derivatives
iteratively in the loop, instead of facing an infinite number of higher-derivative one-body operators 
as indicated in (\ref{field:red:lag}).

Taking into account the trivial rescaling from the relativistic field $\phi$ to the
nonrelativistic field $\Psi$ in (\ref{reparameterize})
(or see Eq.~(\ref{rewrite:propagator})), we obtain the matching equation:
\bqa
  T_0 &=&  4m^2 \, {\cal A}_0 \,,
\label{matching:formula}
\eqa
where $T_0$ is the S-wave partial amplitude projected out of
the $T$-matrix element in the full theory.
This matching formula holds to any loop order.

If we had used (\ref{field:red:lag}) to implement the relativistic corrections,
we should have multiplied the right-hand side of (\ref{matching:formula})
by $(\sqrt{m^2+k^2}/m)^{4/2}= 1+k^2/m^2$, 
to compensate for the modification of the residue of the propagator 
by the field redefinition (\ref{field:redefine}),
in compliance with the Lehmann-Symanzik-Zimmermann reduction formula~\cite{Luke:1997ys}.
Therefore, the matching formula in this scheme  becomes 
\bqa
T_0 &=& 4 \, (m^2+k^2) \, {\cal A}_0^\prime\, .
\label{different:matching}
\eqa
Although  the  tree level amplitude in the relativistic $\phi^4$ theory,
$T_0^{\rm tree}=-\lambda$,  is innocently simple, 
the matching in this scheme becomes unnecessarily involved.

We will employ (\ref{new:matching:Lagr}) to implement the relativistic corrections
throughout this work.
In this scheme,  tree level matching is trivial.
From  the matching formula (\ref{matching:formula}) and
the tree-level EFT amplitude (\ref{tree:amplitude}),
the Wilson coefficients can be simply determined: 
\bqa
  C_0 &=&  {\lambda \over 4m^2} + O(\lambda^2)\,, 
\label{tree:C0}  
  \\   
  C_2 &=&  0 + O(\lambda^2)\,.    
\label{tree:C2}  
\eqa
They can  also be obtained by comparing the Lagrangian 
(\ref{eq:effL}) with (\ref{new:matching:Lagr}).
Since the above derivations  don't assume the specific spacetime
dimensions,
these results hold for the  $\phi^4$ theory living in arbitrary dimensions.
Once one goes beyond the tree level,  however, 
the matching results will generally
depend upon the dimensions.
We will explore the one-loop matching for the $\phi^4$ theory 
in different dimensions in Sec.~\ref{matching:oneloop}.

\section{Can triviality be seen in nonrelativistic limit?}
\label{trivial}

Triviality of the 4D $\phi^4$ theory is intimately connected with the self-consistency
of a non asymptotically-free quantum field theory, and necessarily involves
the consideration of very short distance physics. 
Beg and Furlong's assertion is based on the
nonrelativistic quantum mechanics. 
Since this framework must cease to work when probing the distance 
shorter than the Compton wavelength, 
their claim seems rather counterintuitive-- 
how can a conclusion be trusted when the applicability
of the underlying framework is in trouble?

In this Section, we will examine where the flaw of their 
argument originates. Their strategy is to convert 
this field-theoretic problem to a quantum mechanical one.
They assume that the nonrelativistic limit of $\phi^4$ theory 
can be described by a $\delta$-function potential.
For such a simple potential,
Schr\"{o}dinger equation (or technically more correct, 
Lippmann-Schwinger equation) can be reduced into an 
algebraic equation and  solved analytically.
Nonetheless, the $\delta^3({\bf r})$ potential 
is too singular that one has to regularize the severe ultraviolet
divergences. 
Upon renormalization, they  claim that the two-body scattering 
in a contact potential leads to a trivial renormalized S-matrix, 
thus  establish the triviality of the $\phi^4$ theory 
in the nonrelativistic limit.

Instead of repeating their derivation using Schr\"{o}dinger formalism,
we try to reproduce their renormalization formula 
in the EFT language. 
The main advantage of the field-theoretic method over quantum mechanics 
is that renormalization can be dealt with  in a systematic manner.
For comparison to their results, we first choose cutoff as
the regulator.
As will be elucidated,  it is the problematic way of removing the
ultraviolet power divergences that leads to their incorrect conclusion.
After clarifying the pitfall of renormalization in cutoff scheme, 
we then switch to more convenient dimensional regularization scheme.

\subsection{No triviality if cutoff scheme  used properly}

\begin{figure}[tb]  
\centerline{\epsfysize=2.3truecm  \epsfbox{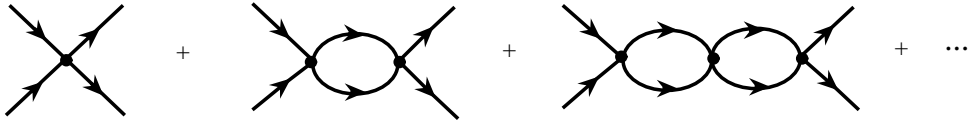}}
\noindent
\caption{The bubble chain  diagrams comprising of $C_0$ vertices.} 
\label{fig:bubbles} 
\vskip .2in
\end{figure}

Since time can only flow forward in the nonrelativistic theory
(Equivalent to say, only one pole is present in the 
nonrelativistic propagator), a gratifying fact is that 
loop calculation in Lagrangian (\ref{eq:effL})
becomes enormously simple.
A further simplification arises from vanishing of many diagrams. 
For example, the self-energy diagrams vanish to all orders,  
so no need for wave-function renormalization.
Particle number conservation reduces the two-body amplitude to the bubble diagrams
as shown in Fig.~\ref{fig:bubbles}. 
This is in sharp contrast with the relativistic $\phi^4$ theory, where
the higher order Feynman diagrams can be arbitrarily complicated.

One remarkable feature is that these bubble chain diagrams 
simply form a geometric series, therefore can be summed analytically.
This should be of no much surprise, as it merely echoes the fact that
one can solve Schr\"{o}dinger equation analytically for
sufficiently simple potential.
For comparison with Beg and Furlong's results,  
it suffices for us to consider the bubble chain with only $C_0$ vertex,
as shown in Fig.~\ref{fig:bubbles}.
Incorporating the higher dimensional operators 
proportional to $C_{2n}$  in the bubble sum
is straightforward, and will be considered in  Section~\ref{matching:oneloop}.
Summing the bubble chain in Fig.~\ref{fig:bubbles}  renders 
\bqa
{\cal A}_0^{\rm sum}  & = &  - C_0^B- C_0^B I_0 C_0^B - 
C_0^B I_0 C_0^B I_0 C_0^B -\cdots 
\label{4d:bubble}
\\
&=& 
-\left[{1\over C_0^B}-I_0\right]^{-1},
\nn
\eqa
where  $C_0^B$ is the bare coupling, and $I_0$ is the one-loop integral 
\bqa
I_0 &= & -{i\over 2} \int{d^4 q \over (2\pi)^4} 
{i \over E+q_0-{\bf q}^2/2m + i\epsilon}\cdot
{i \over E-q_0-{\bf q}^2/2m + i\epsilon}
\nn \\
&= &  -{m\over 2} \int^\Lambda {d^3 {\bf q} \over (2\pi)^3} 
{1 \over {\bf q}^2 -2 m E - i\epsilon}
\nn \\
&=& - {m\over 4\pi^2} \, \Lambda + { m\over 8\pi}\,\sqrt{-2mE-i\epsilon}
+  O(1/\Lambda).
\label{I0:cutoff}
\eqa
We have chosen to work in the C.M. frame, in which each external particle 
carries momentum $k$, and its corresponding kinetic energy $E$ is given in (\ref{rel:dispersion}).
We first carry out the  $q^0$ contour integral
with the aid of Cauchy's theorem.
The remaining 3-momentum integral is ultraviolet-divergent and 
needs regularization.
We impose a momentum cutoff $\Lambda$ to regulate the integral,
and the leading term is linearly divergent.
Recall the one-loop integral in the relativistic $\phi^4$ theory is 
only logarithmically divergent. 
This is because EFT is designed to reproduce the correct low energy property
of the full theory, at a price in distorting the true UV behavior. 
It is generic that EFT always produces  worse UV divergences 
than the full theory.

The second term is finite and imaginary,  
and the residual  $O(1/\Lambda)$ term is small and regulator-dependent,
and can be neglected.
This imaginary term is ordered by the optical theorem,
so has the physical significance and doesn't depend on the regularization prescription.
To respect the positivity of the imaginary part of the amplitude,
we have chosen the convention  $\sqrt{-2mE-i\epsilon}\equiv -i\sqrt{2mE}$.
If we neglect the relativistic correction to the energy-momentum relation, 
this term  equals  $-i k$.

The linear divergence stems from the ultraviolet part of the loop integral,
whose effect cannot be correctly described by the nonrelativistic effective theory, 
therefore renormalization must be invoked.
This divergence can be absorbed into the unknown  bare coupling $C_0^B$, 
by introducing the renormalized coupling $C_0^R$:
\beq
{1\over C_0^R }= {1\over C_0^B} +{m\over 4\pi^2} \, \Lambda \, ,
\label{4d:renormal:cutoff}
\eeq
which is finite and cutoff-independent.
Now the resumed amplitude (\ref{4d:bubble}) can be expressed in terms of
$C_0^R$:
\bqa
{\cal A}_0^{\rm sum} & = & -\left[{1\over C_0^R}+ { i m\over 8\pi}\,k \right]^{-1},
\label{finite:4d:resum}
\eqa
which is also finite and cutoff-independent.

The renormalization relation (\ref{4d:renormal:cutoff}) is the same
as what Beg and Furlong have obtained 
from solving the regularized Lippmann-Schwinger equation,
except an insignificant discrepancy in the coefficient of 
the $\Lambda$ term.

Beg and Furlong argue, for any value of the bare coupling $C_0^B$, 
when one takes the limit $\Lambda \to \infty$,  
the renormalized  $C_0^R$  is forced to vanish,
so is the amplitude in (\ref{finite:4d:resum}).
They then conclude,  the renormalized $\lambda$, related to $C_0^R$  
through (\ref{tree:C0}), must also vanish.
Thus $\phi^4$ theory in the nonrelativistic limit
is said to be trivial.

The key point is that, are we allowed to send the cutoff to infinity 
in (\ref{4d:renormal:cutoff})?
The emergence of power-law UV divergence is a warning sign, 
that this theory cannot hold true at arbitrarily high scale, 
and must break down somewhere.
The cutoff $\Lambda$  should be taken close to the scale where the theory is expected to fail. 
For a realistic system, the cutoff scale is set roughly by the inverse of the range of interaction.
For the $\phi^4$ theory,  the cutoff $\Lambda$ is of order the scalar mass.
If we assume the bare coupling $C_0^B\sim 1/m^2$, 
then the renormalized coupling $C_0^R\sim 1/m^2$,
which is finite. Thus the effective theory (\ref{eq:effL}) 
makes unambiguous and nontrivial predictions.
Generally speaking, taking  $\Lambda \to \infty$ is unacceptable 
for any effective field theory, 
because it pushes the theory way beyond its range of applicability.

Clearly,  Beg and Furlong's assertion is more general than the $\phi^4$ theory,
and applies to any system with the {\it true} $\delta$-function potential.
They try to provide a physical explanation for their assertion--
because two point particles cannot perceive each other 
in a $\delta^3({\bf r})$  potential, therefore no scattering
can occur, so the $S$-matrix is trivial.
This viewpoint, that zero-range interaction leads to
a noninteracting theory,
is further corroborated in Ref.~\cite{Cohen:1996my},
and attributed to the consequence of Friedman's theorem~\cite{Friedman:1972}.

We need inspect, to which extent, the $\delta^3({\bf r})$  potential 
is relevant to the reality? 
There are undoubtedly various realistic systems with 
a repulsive, short-range,  but nontrivial interaction.
These systems can be successfully described by a {\it local} 
nonrelativistic field theory.  If one identifies
local operators with contact  potentials,  then one is puzzled by 
the fact why Beg and Furlong's assertion doesn't apply here.
Of course, the true reason that  local nonrelativistic field theory
can correctly describe the reality is 
not because the interaction is literally {\it contact}, 
but because the range of the interaction is {\it short}
compared with the de Broglie wavelength of particles.

One may still argue, because the scalar particles in  the $\phi^4$ theory
can be viewed as point-like, and this theory is supposed to be valid 
at the distance much shorter than $1/m$,
the two-body interaction may be thought of as truly zero-range.

The very notion of short-range interaction in nonrelativistic 
quantum mechanics deserves some elaboration.
As is  well known, one cannot probe the distance between two nonrelativistic particles
with a resolution better than their Compton wavelengths. 
Otherwise, according to the Uncertainty Principle, 
the energy fluctuation becomes of order $m$, 
the relativistic effects such as pair creation and annihilation 
will invalidate the nonrelativistic quantum mechanics,
and we must resort to the relativistic quantum field theory 
for a correct description.
Therefore, the shortest distance in nonrelativistic quantum mechanics 
which is still meaningful to talk about is the Compton wavelength.
This is another way to say that, in a nonrelativistic problem,
the cutoff should never be taken much bigger than $m$.

In this sense,  $\delta$-function potential should be viewed as an idealized 
mathematical construct, and there is no any physical system possessing 
zero-range interaction.
Therefore, even for the $\phi^4$ theory,  which  supposedly accommodates a contact interaction, 
the interaction range in nonrelativistic scattering
is of order $1/m$, instead of zero.
We will explicitly compute the  S-wave effective range in the $\phi^4$ theory
in Section \ref{matching:oneloop}.

\subsection{Dimensional Regularization}

Although cutoff is a very physical and intuitive regulator,
it is somewhat awkward for practical use.
For instance, when we include the higher-derivative operators  or relativistic corrections,
severer power-law divergences will be confronted. 
In the cutoff scheme, 
the relationship between  renormalized couplings and 
bare couplings in general is very complicated, and 
a particular drawback is that lower-dimensional operators
get renormalized by the higher dimensional ones~\cite{vanKolck:1998bw}.

Physics certainly shouldn't rely on which regulator to use, 
but one judiciously chosen regulator may be more convenient than another.
Dimensional regularization (DR) is the preferred one to 
use practically, especially for the EFT calculation~\cite{Georgi:qn}.
The particular advantages of this scheme 
include that  power-law divergences are automatically subtracted,
the spacetime symmetry (Galilean or Lorentz) is automatically
preserved.
In this scheme, the higher dimensional operators never renormalize
the lower dimensional ones.

For reader's convenience,  we present the master formula 
of DR here,  which will be heavily used in this work:
\bqa
\label{master:DR}
 \int{d^{D-1} {\bf q} \over (2\pi)^{D-1} }  \, 
{({\bf q}^2)^\beta \over ({\bf q}^2 + \Delta)^\alpha}
&=& {1\over (4\pi)^{D-1\over 2} } \,
{\Gamma[\beta+{D-1\over 2}]\,\Gamma[\alpha-\beta-{D-1\over 2}]
\over \Gamma[{D-1\over 2}]\,\Gamma[\alpha] }
\,\Delta^{\beta-\alpha+{D-1\over 2}}\,.
\eqa

Using this formula, we recalculate the one-loop
integral $I_0$ in DR:
\bqa
I_0
&= &  -\left({m\over 2}\right) \, \left(\mu\over 2 \right)^{4-D}\int {d^{D-1} {\bf q} \over (2\pi)^{D-1}} 
\, {1 \over {\bf q}^2 -2 m E - i\epsilon}
\nn \\
&=&  -\left({m\over 2}\right)\, {(\mu/2)^{4-D} \over (4\pi)^{D-1\over 2} } \,
\Gamma \left[{3-D\over 2}\right]
\,\left( -2mE-i\epsilon \right)^{D-3\over 2} \,.
\label{4d:oneloop}
\eqa

It is standard to use {\it minimal subtraction} (MS) in conjunction with DR.
Since this integral doesn't exhibit a $D=4$ pole,
so MS basically does nothing:
\bqa
I_0  &=& { m\over 8\pi}\,\sqrt{-2mE-i\epsilon}\, ,
\label{true:4d:oneloop}
\eqa
which is automatically finite, and  doesn't  depend on the subtraction scale $\mu$.
Note only the finite imaginary part 
in (\ref{I0:cutoff})  survives in the MS scheme.
Plugging (\ref{true:4d:oneloop}) back into (\ref{4d:bubble}),
and replacing $C_0^B$ by $C_0^R$ there,
we reproduce the renormalized amplitude (\ref{finite:4d:resum}) 
which is previously obtained  in the cutoff scheme.
From now on,  unless stated otherwise,
we will always assume that MS (or $\overline{\rm MS}$) is used.
To simplify the notation,  we will suppress the superscript $R$ 
which stands for the renormalized coupling.

It is worth emphasizing that,  throwing away the power divergence
should not be taken for granted.  This is permissible only when 
no delicate cancellation occurs between  two terms in the right-hand side of
(\ref{4d:renormal:cutoff}), so that the renormalized $C_0$ is small.
The $\phi^4$ theory satisfies this criterion.

However, there are a class of interesting systems, 
where the underlying short-distance physics is
both nonperturbative and finely-tuned, so that the two terms in the right side of
(\ref{4d:renormal:cutoff}) conspire to nearly cancel each other, and
results in an unnaturally  large renormalized $C_0$ (or S-wave scattering length).
Two nucleons in S-channel constitutes such an example, 
where the deuteron  manifests as a shallow S-wave bound state.
In this case, cutoff plays an important role in delineating the fine tuning,
and should not be discarded.

If one persists to use DR,  the rule of MS must be altered  correspondingly 
to limn the effects of fine tuning.
An ingenious scheme,  {\it power divergence subtraction}  (PDS), 
has been  introduced  for better describing such a finely-tuned system~\cite{Kaplan:1998tg}. 
It is a generalization to MS,
and  we will encounter it in the next Section.

\section{One-Loop Matching and Bubble Chain Sum}
\label{matching:oneloop}

In this Section, we will present a systematic study of short-range force in various spacetime dimensions: 
two, three and four.
This Section is divided into three parts,  each of which
devotes a detailed discussion to each case.
Among them, the 3D case is especially interesting, where the renormalization group 
technique can be fruitfully applied.

In each part, we first take the $\phi^4$ theory as a concrete example of short-range interaction,
and match it onto the nonrelativistic EFT at one-loop level.
This is a simple, but ideal place to illustrate 
the generic features of perturbative matching,  {\it e.g.} 
cancellation of non-analytic terms and infrared divergences, etc.

We then proceed to explore the nonrelativistic EFT sector.
It is shown that the bubble chain diagrams can be summed analytically,
with higher-dimensional operators and relativistic effects fully incorporated.
The resumed amplitude can be framed in a compact form.
From this exact result, we can gain  nonperturbative understanding 
of short-range interactions in the two-body sector.

\begin{figure}[tb]
\centerline{\epsfysize=5.0truecm \epsfbox{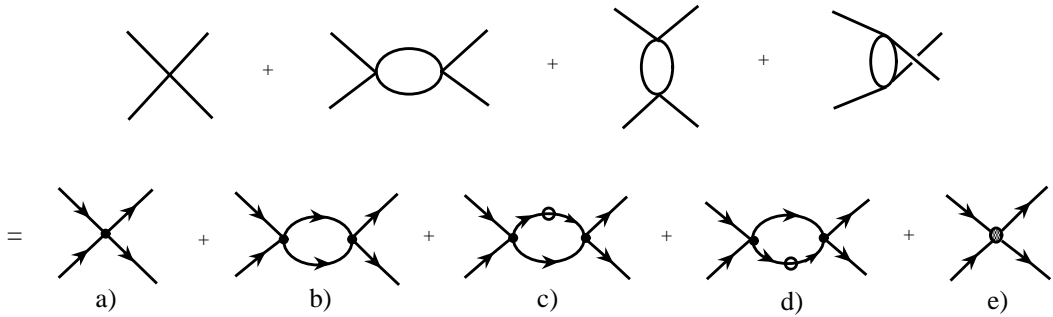}}
\noindent
\caption{Diagrams for one-loop matching. The top row represents the diagrams
in the full theory, and the bottom row represents those in the effective theory.
The $C_0$ vertex is denoted by a solid dot, $C_2$ vertex depicted by a shaded circle, and the two-legged relativistic vertex 
is represented by an open circle. 
Counterterms are not explicitly shown. }
\label{fig:match} 
\vskip .2in
\end{figure}

Let us now sketch how to carry out the one-loop  matching.
The three one-loop diagrams in the relativistic $\phi^4$ theory
are shown in the top row of Fig.~\ref{fig:match}.
The $s$-channel diagram contains an imaginary part, 
the other two, the $t$- and $u$-channel diagrams are real.

After Feynman parameterization and integrating over loop momentum,
the $T$-matrix in the $D$-dimensional $\phi^4$ theory
is~\cite{Peskin:ev}
\bqa
 T^{\rm  1-loop} &=&  -\lambda +{\lambda^2 \over 2 \,m^{4-D}} {\Gamma[2-D/2] \over (4\pi)^{D/2} } 
\, \int_0^1  dx \, \left\{ \left[ 1-x(1-x) s/ m^2 -i\epsilon \right]^{ D/2-2} \right.
\nn \\
 & & \qquad \qquad + \left. (s \rightarrow t) + (s\rightarrow u)  \, \right\} - \delta \lambda\,,
\label{full:amplit:allpartial:waves} 
\eqa
where $\delta \lambda$ is the counterterm.
The Mandelstam variables $s$, $t$ and $u$ 
are defined
by $s\equiv(k_1+k_2)^2$, $t\equiv(k_1-k_1^\prime)^2$,
$u\equiv(k_1-k_2^\prime)^2$.
In the C.M. frame, these variables become
$s=4(m^2+k^2)$, $t= -2 k^2 (1-\cos\theta)$ and $u= -2 k^2 (1+\cos\theta)$,
where $\theta$ is the angle between ${\bf k}_1$ and ${\bf k}_1^\prime$.

The amplitude in (\ref{full:amplit:allpartial:waves}) 
is  the superposition of all the partial waves of even angular momentum,
and only the S-wave amplitude $T_0$ needs to be projected out.
Bose statistics implies that the occurrences of $t$ and $u$ 
are symmetric. Because $|t|,\,|u|\ll m^2$,  we can expand the amplitude in terms of
$t/m^2$, $u/m^2$ and only keep the linear combination of 
these two variables:
\bqa
t+u & = & 4m^2-s = -4 k^2\,,
\label{linear:comb:tu}
\eqa
which doesn't have angular dependence.
This guarantees that correct $T_0$ is projected out
up to $O(k^2)$, because the P-wave amplitude is absent
and the D-wave contribution starts at order $k^4$.

We also need consider the diagrams in the EFT sector, 
as shown in the bottom row of  Fig.~\ref{fig:match}.
The S-wave amplitude to one-loop order is
\bqa
\label{match:eft:formula}
 {\cal A}_0^{\rm 1-loop}  &=& 
  -C_0 - C_0 (I_0+\tilde{I_0})\,C_0 - C_2\, k^2-
  \delta C_0 - \delta C_2\, k^2.
\eqa
Here $I_0$ stands for the one loop integral corresponding  to Fig.~\ref{fig:match}b),
and $\tilde{I_0}$,  denoted by Fig.~\ref{fig:match}c) and d),
represents the one loop integral with one insertion of the relativistic vertex.
The tree level contribution from  $C_2$  vertex is shown in Fig.~\ref{fig:match}e). 
Counterterms are also implied when the subtraction scheme is specified.

Requesting the S-wave amplitude calculated in the full theory and the EFT 
to be equal,
we can deduce the values of $C_0$, $C_2$, \ldots, respectively.

\subsection{Two dimensions}
\label{2d:phi4:subsection}

Everyone is familiar with the one-dimensional $\delta$-function potential.  
The transmission and reflection probability can be easily obtained
by solving  Schr\"{o}dinger equation. 
For the attractive $\delta$-function potential, 
there is also a bound state solution.
However,  it is difficult to apply the Schr\"{o}dinger formalism
to more general situations, 
{\it e.g.}, inclusion of $\delta^{\prime\prime}(x)$ potential, and
incorporating  relativistic corrections, and so on.
As we will see,  these questions are best addressed
by the field-theoretic approach.

\subsubsection{Matching of 2D $\phi^4$ theory}

The two-dimensional $\phi^4$ theory has a coupling $\lambda$ of mass dimension two,
so is super-renormalizable. 
Since no ultraviolet divergences emerge,  renormalization is not required.

Substituting $D=2$ into (\ref{full:amplit:allpartial:waves}),
we work out the following integral associated with the $s$-channel diagram:
\bqa
\int_0^1 \: dx \, [1- x(1-x)\,s/m^2-i\epsilon ]^{-1} &=&
{2m\over \sqrt{s}} \, {m\over k}\, \left(
\tanh^{-1} \beta + {i\pi \over 2}  \right) 
\nn \\
&\approx& 1  -{2 k^2\over 3 m^2} + {i \pi m \over 2 k} 
\left(1-{k^2\over 2m^2} \right) \,,
\label{2D:s-channel}
\eqa
where $\beta=\sqrt{1-4m^2/s}$  is the velocity 
in the C.M. frame.
We have expanded both the real and imaginary part 
to the relative order $k^2$.

The leading imaginary part is singular at small momentum,
which  can  be understood from the optical theorem,
due to the too limited phase space.
This should be viewed as an artifact of the perturbative expansion,
and the sensible answer will be obtained only if the singular terms
are summed to all orders.

For the $t$- and $u$-channel diagrams, the corresponding integrals are 
\bqa
\int_0^1 \: dx \,\left\{ [1- x(1-x)\,t/m^2 ]^{-1} + (t\rightarrow u) \right\}
&\approx&
2+{t+u\over 6m^2} = 2 - {2k^2\over 3m^2}.
\label{2D:tu-channel}
\eqa
We first expand the integrand in $t/m^2$ and $u/m^2$ to the first order, 
combine them in compliance with (\ref{linear:comb:tu}),  then carry out the integration over $x$.
Note the result is convergent at small $k$.

Merging (\ref{2D:s-channel}) and (\ref{2D:tu-channel}),
we obtain the amplitude to one loop order\footnote{
Since angular momentum cannot be defined in one spatial dimension,
so partial wave expansion loses its meaning.  We therefore drop the subscript 0, 
which stands for the S-wave.}:
\bqa
 T^{\rm 1-loop}  &=&  -\lambda +{\lambda^2 \over 8 \pi m^2} \,
\left[ 3 + {i\pi m\over 2 k} 
\left( 1-{k^2\over 2m^2} \right) - {4 k^2\over 3 m^2} \right]\,.
\label{2D:FULL:AMPL}
\eqa

We  now  move on  to the EFT sector. Since (\ref{4d:oneloop}) doesn't exhibit 
$D=2$ pole,  we can directly substitute $D=2$ into this equation, 
and find that Fig.~\ref{fig:match}b) equals
\bqa
I_0 &= &  -{m \over 4 \sqrt{-2mE-i\epsilon} }
\approx - {im\over 4 k}\,\left(1+{k^2\over 8m^2}\right),
\label{2d:oneloop:full}
\eqa
where we have retained the first-order relativistic correction 
in $E$, in conformity with (\ref{rel:dispersion}).

It is straightforward to evaluate the one-loop diagrams 
with  one relativistic vertex insertion, Fig.~\ref{fig:match}c) and d):
\bqa
\tilde{I_0} &=& -i\,\left({\mu\over 2 }\right)^{2-D}\int{d^D q \over (2\pi)^D}
\left( { i \over E+q_0-{\bf q}^2/2m + i\epsilon} \right)^2
 \cdot {i(E+q^0)^2\over 2m} \cdot
{i \over E-q_0-{\bf q}^2/2m + i\epsilon}
\nn  \\
 &\approx &  -{1\over 8m}\, \left({\mu\over 2 }\right)^{2-D} \int{d^{D-1} {\bf q} 
 \over (2\pi)^{D-1}}\, 
{ 3k^4-2k^2 {\bf q}^2 \over ({\bf q}^2 -k^2 - i\epsilon)^2} 
\nn \\
 &=&  { (\mu/2)^{2-D} \over (4\pi)^{D-1\over 2}} \, 
 \left( \, 3 \, \Gamma\left[{5-D\over 2}\right] +
(D-1) \,\Gamma \left[{3-D\over 2} \right] \,\right) \, 
\left({k^2\over 8m} \right) \, (-k^2-i\epsilon)^{D-3\over 2}   
\, .
\label{D:oneloop:rel}
\eqa
After integrating out the variable $q^0$,
we simplify the expression little bit
by employing  the fact that scaleless integrals vanish in DR.
We have also replaced $2mE$ by $k^2$, since the induced 
error is of relative order $k^4$.

This formula  doesn't display a $D=2$ pole either, so we simply replace $D$ by 2
everywhere, and the result is
\bqa
\tilde{I_0} &=&  {im \over 4k}\left({5 k^2 \over 8m^2}\right).
\label{2d:oneloop:rel}
\eqa

According to (\ref{match:eft:formula}), 
we obtain the one-loop EFT amplitude 
by combining (\ref{2d:oneloop:full})  and  (\ref{2d:oneloop:rel}):
\bqa
 {\cal A}^{\rm 1-loop} &=&  -C_0 + i\,C_0^2\,{m\over 4 k} \,
 \left( 1-{k^2\over 2m^2} \right) - C_2\,k^2.
\label{2D:EFT:AMPL}
\eqa

Comparing (\ref{2D:FULL:AMPL}) and (\ref{2D:EFT:AMPL})
through (\ref{matching:formula}),
we see that the full theory and the EFT  share the same non-analytic (imaginary) terms.
Note the structures of relativistic corrections are the same in both sectors.
All of these are ensured by the general principles of EFT.
Consequently, we obtain the Wilson coefficients:
\bqa
 C_0 &=& {\lambda \over 4 m^2} - {3\over 2\pi} \left( {\lambda \over 4 m^2}\right)^2 
 + O(\lambda^3)\,, \\
 C_2 & = &  {2\over 3\pi m^2} \left( {\lambda \over 4 m^2}\right)^2
 + O(\lambda^3) \,  . 
\eqa

Through the one-loop matching, $C_0$ receives an $O(\lambda^2)$  correction,
and  a  nonzero coefficient is generated for $C_2$.
Tracing back to the relativistic $\phi^4$ theory calculation, 
we can identify how  those three different channels contribute in the matching.
Because the $t$-, $u$-channel processes don't have counterparts in the EFT sector,
their effects are entirely  encoded in the Wilson coefficients,
and not responsible to the relativistic corrections\footnote{Note {\it t},
{\it u} are simple polynomials of momentum $k$. In contrast, the $s$-channel
integral contains the factor of $\sqrt{s}$, which is an infinite power series in $k$.}.
In contrast,  the $s$-channel diagram not only contains dynamical effects, but 
also contains the purely kinematic effects-- the relativistic corrections.
At one loop order, the latter only influences its non-analytic (imaginary) part.

\subsubsection{Bubble chain sum in 2D EFT}

The  $k\to 0$  singularity in the one-loop amplitude (\ref{2D:EFT:AMPL}) 
implies that fixed-order perturbative expansion is not reliable.
Since each loop contributes a factor of $im/4k$,  higher order diagrams 
become more singular.
To remedy this nonphysical singularity, it is mandatory to sum 
the infrared-divergent terms to all orders.

This is an almost intractable task in the relativistic $\phi^4$ theory,
because any diagram, no matter how complicated, so long as 
containing one $s$-channel subdiagram,  needs to be included.
However,  this problem becomes rather transparent in the nonrelativistic EFT.
Let us first consider summing the most singular series-- the bubble chain comprising of $C_0$ vertices only.
Analogous to (\ref{4d:bubble}), this can be easily accomplished:
\beq
{\cal A}^{\rm sum} 
= -\left[{1\over C_0} +
 {im\over 4 k} \right]^{-1} \,.
\label{bubble:2d:C0only}
\eeq
We immediately see that, the infrared singularity confronted in 
the fixed-order perturbative expansion is now removed.  
As $k \to 0$, the resumed amplitude 
vanishes as $4ik/m$,  not dependent of $C_0$ at all.
This result can be obtained from solving Schr\"{o}dinger equation,
and can be understood from that a particle carrying very long wavelength cannot penetrate
through the  one-dimensional infinitely high barrier.

This equation also encodes another important nonperturbative information.
The pole of the amplitude located at the positive imaginary momentum, 
$k=i\kappa$ ($\kappa>0$), signals a bound state with binding energy
\beq
E_B \equiv 2 \, (\sqrt{m^2 - \kappa^2}-m) =
- {\kappa^2 \over m} -\cdots .
\label{binding:energy}
\eeq

One easily finds the location of the pole in (\ref{bubble:2d:C0only}), 
$\kappa=-m\, C_0/4$.
In order to have a positive $\kappa$, we must require $C_0<0$,   
which corresponds to an attractive $\delta$-function potential. 
The binding energy is then about $-m C_0^2 /16$.
This is nothing but the familiar result for a particle with the reduced
mass $m/2$ in the $-{|C_0|\over 2}\delta(x)$ potential.
Clearly, the weaker the coupling, the shallower
the bound state is.

Including those higher-derivative operators in the bubble diagrams 
ameliorates the infrared behavior,
but not sufficient to justify a fixed-order calculation,
since the singularities will emerge when enough bubble diagrams are retained.
To obtain  physically sensible result,
we must also sum  bubble chains containing these operators.

The loop integral needed to evaluate these bubble diagrams is
\bqa
I_n  &= & -\left( {i\over 2} \right) \, \left({\mu\over 2}\right)^{2-D}  
\int{d^D q \over (2\pi)^D} \, {\bf q}^{2n} \,
{i \over E+q_0-{\bf q}^2/2m + i\epsilon}\cdot
{i \over E-q_0-{\bf q}^2/2m + i\epsilon}
\nn \\
&= &  - \left( {m\over 2} \right) \, \left({\mu\over 2}\right)^{2-D}
\int {d^{D-1} {\bf q} \over (2\pi)^{D-1}} 
\, { {\bf q}^{2n} \over {\bf q}^2 -2 m E - i\epsilon}
\nn \\
&=&  -\left({m\over 2}\right)\, {(\mu/2)^{2-D} \over (4\pi)^{D-1\over 2} } \,
\Gamma \left[{3-D\over 2}\right]
\,\left( -2mE-i\epsilon \right)^{D-3\over 2} \, (2mE)^n \,.
\label{good:loop:feature}
\eqa
Therefore,  the following relation holds in any dimensions: 
\bqa
I_n &=&   I_0\, \left( 2mE \right)^n  \approx I_0\,k^{2n}\, ,
\eqa
where we neglect the relativistic effect in the last equality.

This relation allows that the  factors of $q$
inside the loop get converted into factors of the external momentum $k$.
To appreciate this gratifying feature,
let us consider one explicit example. 
First consider the one loop diagram with $C_0$ and $C_2$ as
its vertices.
It contributes to the amplitude with 
\bqa
& &   C_0\, C_2 \, \left( {m\over 2} \right) \,\left({\mu\over 2 }\right)^{2-D}\int {d^{D-1} {\bf q} \over (2\pi)^{D-1}} 
\, { k^2 + {\bf q}^2 \over {\bf q}^2 -2 m E - i\epsilon}
\nn \\
&=&  -C_0 \, C_2 \,(k^2 I_0 + I_1) \approx - 2\, C_0 C_2\,I_0\,k^2  \,.
\label{c0c2:2d:oneloop}
\eqa
Similarly, the one loop diagram with two $C_2$ vertices contributes
\bqa
& &    \left( {C_2\over 2} \right)^2 \, \left( {m\over 2} \right) \,
\left({\mu\over 2 }\right)^{2-D}\int {d^{D-1} {\bf q} \over (2\pi)^{D-1}} 
\, { (k^2 + {\bf q}^2)^2 \over {\bf q}^2 -2 m E - i\epsilon}
\nn \\
&=&  -   \left( {C_2\over 2} \right)^2 \,(k^4 I_0 + 2 k^2 I_1 + I_2) \approx - C_2^2 \,I_0\,k^4 \,.
\label{c2c2:2d:oneloop}
\eqa
Recall the one loop diagram with two $C_0$ vertices contributes to the amplitude with $-C_0^2 I_0$.
These three terms can be combined into a simple form, $-(C_0+C_2 k^2)^2\,I_0$.

This suggests that,  we can lump all the two-body S-wave operators together, 
and treat them as a single effective operator. 
Consequently, we replace each internal $C_0$ vertex in the bubble chain 
depicted in Fig.~\ref{fig:bubbles}
by an effective vertex $-\sum C_{2n} k^{2n}$~\cite{Kaplan:1998tg}.
It can be checked that combinatorics is correctly taken into account.
The full bubble chain sum thus gives
\bqa
{\cal A}^{\rm sum} &=& -\left[{1\over C_0+C_2\,k^2+\cdots } +
 {im\over 4 k} \right]^{-1}.
\label{bubble:2d:all}
\eqa
From this equation, one can infer that a bubble chain containing $n$ $C_0$ vertices
and one $C_2$ vertex contributes with
$(-1)^{n+1} (n+1) C_0^n C_2 (im/4k)^n k^2$.  Taking $n=1$, one reproduces the
one loop result in (\ref{c0c2:2d:oneloop}).

When the momentum gets small,
the higher-dimensional terms become negligible relative to the $C_0$ term.
Ultimately in the $k\to 0$ limit, the amplitude is again
governed by the reciprocal of the imaginary factor $I_0$.

Including higher-dimensional terms will shift the bound state pole.
From this equation, one finds that the bound state pole moves to
\bqa
\kappa &=&  {2\, \left(1-\sqrt{1 - |C_0|\, C_2  m^2/4}\; \right)\over m\,C_2}
\nn \\
& \approx &  { m |C_0| \over 4} 
\left[ 1+ {|C_0|\, C_2 m^2\over 16}\right] \, .
\label{ppole:kappa:2d}
\eqa
In order for $\kappa$ to have a real solution, one needs impose  $C_2<4/(m^2\,|C_0|)$.
For small enough $C_2$,
the corresponding binding energy $E_B\approx -{m C_0^2\over 16} [1+ |C_0| C_2 m^2/8]$.

Thus far, the relativistic corrections have been omitted in the resumed amplitude.
Intuitively, relativistic effects ought to be negligible at small $k$.
However, because of the rising of infrared singularity,  we are not allowed to ignore them {\it a priori}.
To elucidate this point, let us consider a $n$-loop bubble chain consisting entirely of $C_0$ vertices, 
but with the first-order relativistic correction included.
It contributes with $(im/4k)^n k^2$,  the same order as the bubble chain containing
$n$ $C_0$ vertices and one $C_2$ vertex.
At $n =3$, infrared divergence arises and keep deteriorating as $n$ increases.
Therefore, to have a physically meaningful answer, 
it is mandatory to sum all the singular terms induced by the relativistic corrections.

At first sight, including relativistic effects in the resumed amplitude
is an impossible task, because the pattern of relativistic
corrections seems too complicated and random to identify.
However,  this is just a disguise,
and a thorough  scrutiny  shows that it is in fact feasible 
to fully incorporate the Lorentz symmetry.
In any event, 
a correct resummation formula must first reproduce the one-loop amplitude
(\ref{2D:EFT:AMPL}), where the first-order relativistic correction is included. 
It is then natural to propose the following formula: 
\beq
{\cal A}^{\rm sum} = -\left[{1\over C_0+C_2\,k^2+\cdots } +
 {im\over 4 k} \, \left( 1-{k^2\over 2m^2} +\cdots \right) \right]^{-1}.
\label{bubble:2d:rel}
\eeq
One can deduce from this equation, that the coefficient of $(im/4k)^n k^2$,
which  arise from the $n$-loop $C_0$ bubble chain implementing
the first-order relativistic correction,  is  $(-1)^n n C_0^{n+1}$.
This can be easily confirmed by direct calculation.
We also verify this resummation formula by computing the first-order relativistic correction
to the $n$-loop bubble diagram  containing $n$ $C_0$ vertices and one $C_2$ vertex.

To complete this resummation formula, we need know the successive terms
in the parenthesis in (\ref{bubble:2d:rel}).
For instance, to pinpoint the $O(k^4)$ term,  we need expand $I_0$ to incorporate the 
second-order relativistic correction,
include the first-order relativistic correction in $\tilde{I_0}$,
and also calculate the one loop diagram with two insertions of the relativistic vertex.

Fortunately, we actually don't need  go through this computation, 
thanks to a shortcut provided by the $\phi^4$ theory.
Relativistic correction, as a solely kinematic effect which only depends on the spacetime symmetry, 
must be identical in the full theory and the EFT calculation.
Therefore, we can directly recognize the pattern of relativistic corrections from the
one-loop $s$-channel integral in the $\phi^4$ theory.
Inspecting the imaginary part in (\ref{2D:s-channel}),  
one finds that the full series
in the parenthesis in (\ref{bubble:2d:rel}) is nothing but
\beq
\gamma^{-1} = {2m \over \sqrt{s}} = 1 - {k^2 \over 2 m^2} + {3 k^4 \over 8 m^4} -\cdots.
\label{dilation}
\eeq
where $\gamma \equiv 1/\sqrt{1-\beta^2}$ is the familiar dilation factor.

This resummation formula can be confirmed  by  miscellaneous straightforward calculations.
This is an amazing result-- though the intermediate stage of computing relativistic corrections 
looks rather involved and desultory,  the final results obey a very simple pattern.

Note when the relativistic corrections are incorporated, 
this resumed amplitude is still well behaved at small $k$.
Specifically, as $k\to 0$,  one finds ${\cal A}^{\rm sum}\approx  4 i \gamma\,k/m $.
Now it is safe to conclude  {\it a posteriori},
that relativistic effects are indeed unimportant at small momentum.

Finally let us investigate the impact of relativistic effects on the bound state pole.
Inspecting (\ref{bubble:2d:rel}), we find that the pole 
shifts  from (\ref{ppole:kappa:2d}) by an additional amount of $\Delta\kappa\approx m C_0^3/64$. 
The corresponding binding energy is then
\bqa
E_B &= & - {\kappa^2 \over m} -  {\kappa^4 \over 4 m^3} -\cdots 
\approx
-{m \, C_0^2 \over 16} \left[ 1 +  {|C_0|\, C_2 m^2 \over 8} 
- {7 \, C_0^2 \over 64} \right] \,,
\eqa
where the net relativistic effect is encoded in the third term,
which slightly reduces the binding energy.
Note its  size is comparable to the second term, when $C_2$  respects the bound
imposed earlier.
Apparently, this result can not be easily obtained  from solving Schr\"{o}dinger equation 
with the potential $-{|C_0|\over 2}\delta(x) - { C_2\over 4}\delta^{\prime\prime}(x)$.

\subsection{Three dimensions}
\label{3d:phi4:subsection}

The nonrelativistic planar system displays rich physics.
For example,  point particles on a plane coupled with the Chern-Simmons gauge field,
can be used to formulate the Aharonov-Bohm effect and Fractional Quantum
Hall effect in a field-theoretic framework~\cite{Hagen:rp}.

We will focus on the simplest case, where the external sources are absent and only
the short-range interactions  among particles themselves are present. 
Even this case is quite nontrivial.
The fullest discussion on this topic by far
which utilizes the field-theoretic language,  is given by Bergman~\cite{Bergman:1991hf}.
In the following, we will expand his results 
to  incorporate the effects of higher-derivative operators and relativity.  
As will be seen, the latter plays an important role in influencing the renormalization
group flow of the former.

\subsubsection{Matching of 3D $\phi^4$ theory}

In the 3-dimensional $\phi^4$ theory 
the coupling $\lambda$ has mass dimension one.
This theory is super-renormalizable, and ultraviolet finite at one loop.

Substituting $D=3$ into (\ref{full:amplit:allpartial:waves}), 
we carry out the $s$-channel integral and expand it: 
\bqa
\int_0^1 \, dx \, [1- x(1-x)\,s/m^2-i\epsilon]^{-{1\over 2}} &=&  
{m\over \sqrt{s}}\,\left(
\ln \left[ {\sqrt{s} +2m \over \sqrt{s}-2m } \right ] +i\pi \right )
\nn \\
&\approx& \left(1-{k^2\over 2m^2} \right)\left[ \ln\left({2m\over k}\right) +
{i\pi \over 2} \right]+{k^2\over 4m^2}\,.
\label{3d:s-channel:integr}
\eqa
This integral is logarithmically divergent as $k\to 0$,
though milder than the linear divergence encountered in 2D.
As emphasized previously, this infrared singularity  is a symptom
that the fixed-order perturbation series is untrustworthy
at small momentum.

The $t$- and $u$-channel integrals can be performed
by exploiting the same trick as in (\ref{2D:tu-channel}):
\bqa
\int_0^1 \: dx \, \left\{ [1- x(1-x)\,t/m^2]^{-{1\over2}} + (t\rightarrow u) \right\}
 &\approx&
2+{t+u\over 12 m^2} = 2 - {k^2\over 3m^2},
\eqa
which is finite in the $k\to 0$ limit.

Thus, to the one-loop order, the S-wave amplitude in the $\phi^4$ theory is
\bqa
 T_0^{\rm 1-loop} &=&  -\lambda +{\lambda^2 \over 16 \pi m} \,
\left[ 2+\left(1-{k^2\over 2 m^2} \right) \,
\left[ \ln\left({2m\over k}\right) + {i\pi \over 2} \right]
-{k^2\over 12 m^2} \right]\,. 
\label{T0:3D:ampl}
\eqa
This expression has previously been obtained in Ref.~\cite{Gomes:1996px}.

Loop integrals in the 3D EFT exhibit a novel feature,
{\it e.g.} emergences of logarithmic UV divergences
(look at the $D=3$ pole in (\ref{4d:oneloop}), (\ref{D:oneloop:rel}),
(\ref{good:loop:feature})).
The log divergence implies that different momentum regions
are coupled, so the coefficients of the logarithms are
regularization-scheme independent.

The one-loop integral corresponding to Fig.~\ref{fig:match}b) is
\bqa
I_0 &=& 
 -{m\over 2}  \left({e^{\gamma} \mu^2 \over 4\pi}\right)^{\epsilon}
 \int {d^{2-2\epsilon} {\bf q} \over (2\pi)^{2-2\epsilon}} \,
{1 \over {\bf q}^2 -2 m E - i\epsilon}
\nn \\
&=&  - {m \over 8\pi}
\,\left[{1\over\epsilon}+ \ln\mu^2-\ln(-2mE-i\epsilon)  \right],
\label{3d:oneloop}
\eqa  
where we have rewritten $D= 3-2\epsilon$,  
and $\gamma=0.5772\cdots$ is the Euler's constant.
We adopt  $\overline{\rm MS}$, by replacing $\mu^2\to e^{\gamma}\mu^2/4\pi$.
Expanding $2mE$ to include the first-order 
relativistic correction, we obtain
\bqa
I_0 &\approx &  -{m \over 8\pi}\,
\,\left[{1\over\epsilon}+ 2\ln\left({\mu\over k}\right) +i\pi 
+ {k^2 \over 4m^2} \right].
\label{3d:new:oneloop}
\eqa

Fig.~\ref{fig:match}c) and d) can be easily deduced from (\ref{D:oneloop:rel}):
\bqa
\tilde{I_0} &=& 
-{1\over 8m}
 \left({e^{\gamma} \mu^2 \over 4\pi}\right)^{\epsilon}
 \int{d^{2-2\epsilon} {\bf q} \over (2\pi)^{2-2\epsilon}}\, 
{ 3 k^4- 2 k^2 {\bf q}^2 \over ({\bf q}^2 -k^2 - i\epsilon)^2} 
\nn \\
&=&  {m \over 8\pi}\, \left({k^2 \over 2m^2}\right)
\,\left[{1\over\epsilon}+2 \ln\left({\mu\over k}\right) +i\pi+{1\over 2}\right]\, .
\label{3d:rel:oneloop}
\eqa
The appearance of $k^2/\epsilon$ pole implies
that $C_2$ is renormalized by $C_0$ through the relativistic correction.

Combining (\ref{3d:new:oneloop}) and (\ref{3d:rel:oneloop}), 
we obtain the S-wave amplitude in the EFT sector:
\bqa
 {\cal A}_0^{\rm 1-loop} &=&  -C_0 + C_0^2 \left( {m\over 4\pi} \right) \,
\left(1- {k^2\over 2 m^2}\right) \left[ \ln \left({\mu \over k}\right)
+ {i\pi \over 2} \right] -C_2 k^2\,, 
\label{A0:3D:1loop}
\eqa
where we have introduced the following counterterms
to absorb the divergences:
\bqa
\delta C_0 &=&  {m\over 8\pi}\,{C_0^2\over \epsilon},
\label{counterterm:C0}
\\
\delta C_2 &=&  -{1\over 2 (8\pi)m}\,{C_0^2\over \epsilon}.
\label{counterterm:C2}
\eqa

Note when $I_0$ and $\tilde{I_0}$ are added together, 
those real analytic $O(k^2)$  pieces exactly cancel. 
It is  generic that the linear combination $\ln(\mu/k)+i\pi/2$
constitutes the only allowed loop factor  accompanying powers of $k$ in the amplitude.
Because of this cancellation,  whether including  relativistic corrections or not 
will not mess up with the to-be-determined 
analytic part of the $C_2$ coefficient.

Comparing (\ref{A0:3D:1loop}) with (\ref{T0:3D:ampl}) via (\ref{matching:formula}),
it is easy to check that the non-analytic terms of the form $\ln k$ and $k^2 \ln k$  
are exactly identical in the full theory and the EFT. This is the designed feature 
of matching~\cite{Georgi:qn}. Of course, the structure of 
the relativistic corrections must be the same.

Consequently, we can determine the Wilson coefficients:
\bqa
 C_0(\mu) &=& {\lambda \over 4 m^2} - \left( {\lambda \over 4 m^2}\right)^2 
 \left( {m\over 4\pi} \right) \,
\left[2 + \ln \left({2m \over \mu}\right)  \right]+O(\lambda^3)\,, 
\label{3D:C0:oneloop:mat} 
\\
C_2(\mu) &= & \left( {\lambda \over 4 m^2}\right)^2  
\left( {1\over 8\pi m} \right) \, 
\left[{1\over 6} + \ln \left({2m \over \mu}\right)  \right]+O(\lambda^3)\,.
\label{3D:C2:oneloop:mat}
\eqa
Both of the coefficients are $\ln\mu$ dependent.

We notice that, the authors of Ref.~\cite{Gomes:1996px} 
don't realize that the subleading non-analytic term,  $k^2 \ln(2im/k)$
in (\ref{T0:3D:ampl}), should be identified with the relativistic correction.
They introduce two  {\it ad hoc}  four-boson operators at second order of $\nabla$,
and adjust their coefficients to reproduce this term 
by considering the one loop diagram with these operators and $C_0$ as vertices.
It should be reminded that, relativistic corrections represent solely kinematic effects,
and to reproduce them don't need involve any unknown  parameters.
In the one-loop matching considered above, 
knowing the tree-level value of $C_0$ suffices to give the correct answer.

In order not to spoil the perturbative matching,
one should  choose the matching scale $\mu$
around $2m$,  to avoid large logarithms.
From (\ref{3D:C0:oneloop:mat}) and (\ref{3D:C2:oneloop:mat}),
one sees that $C_0$ decreases and $C_2$ increases as $\mu$ decreases.
The logarithms grow as $\mu$ declines, 
and ultimately it is more secure to call for the renormalization group (RG) equation
to sum these large logs.

\subsubsection{3D EFT and renormalization group}

Having considered the 3D $\phi^4$ theory as a specific example 
to illustrate the matching procedure,  we now turn to general 
discussions on the planar system accommodating short-range interactions.
We will derive the exact RG equations for the Wilson coefficients, 
and also present an exact nonperturbative expression for the S-wave scattering amplitude.

We start with deriving the RG equation for $C_0$.
The bare $C_0$ can be expressed as
\bqa
C_0^B &=& \mu^{2\epsilon} \left[C_0+ {m\over 8\pi} {C_0^2\over \epsilon} +
\left({m\over 8\pi}\right)^2 {C_0^3\over \epsilon^2} +\cdots \right]
\nn \\
&=& {\mu^{2\epsilon}  C_0 \over 1-{m\over 8\pi} {C_0\over \epsilon} }\,.
\label{bareC0:renormC0}
\eqa
The leading-order counterterm is given in (\ref{counterterm:C0}),
and the succesive ones simply form a geometric series. 
The absence of subleading poles at any loop order 
indicates that an exact RG equation for $C_0$ can be deduced.
Acting $\mu \, d/d\mu$ to the above equation, 
applying the chain rule, one obtains the $\beta$ function for $C_0$:
\bqa
\beta(C_0,\epsilon) &=& -{2\epsilon \,C_0^B(C_0,\epsilon) 
\over \partial  C_0^B/\partial C_0|_\epsilon }
\nn \\
&=& -2\epsilon \, C_0 + {m\over 4\pi}\, C_0^2\,.
\label{beta:function:C0}
\eqa
Since the right-hand side is positive,  the free coupling limit 
is the infrared fixed point.
The solution  of this RG equation is~\cite{Bergman:1991hf}
\bqa
C_0(\mu) &=&  \left[{1\over C_0(\Lambda)}+{m\over 4\pi} 
\ln\left({\Lambda\over \mu}\right) \right]^{-1}\, ,
\label{solution:RG:C0}
\eqa
where  $\Lambda$, which characterizes the breakdown scale of the 
nonrelativistic EFT,  together with $C_0(\Lambda)$  comprise the boundary conditions.
For the nonrelativistic $\phi^4$ theory, we may choose
$\Lambda=2m$, and $C_0(2m)$ can be read off from (\ref{3D:C0:oneloop:mat}).

In the $k \to 0$ limit,  fixed-order bubble diagrams with $C_0$ vertex
suffers from logarithmic singularity.
One hopes that after a nonperturbative bubble sum,  
the infrared behavior will ameliorate. 
Analogous to (\ref{4d:bubble}) , the bubble diagrams
consisting entirely of $C_0$ vertices can be easily summed:
\bqa
{\cal A}_0^{\rm sum}  & = &  
-\left[{1\over C_0(\mu)}+{m\over 4\pi} 
\left[ \ln\left({\mu\over k}\right) + {i\pi\over 2} \right]
\right]^{-1} .
\label{3d:bubble}
\eqa
Requesting it to be $\mu$ independent,
one quickly reproduces the RG solution (\ref{solution:RG:C0}).
To avoid the large logarithm,
the optimal renormalization scale $\mu$ should be chosen around $k$.
From (\ref{solution:RG:C0}), we see that the effective coupling $C_0(\mu)$  
vanishes inverse logarithmically as $\mu$  approaches zero.
As $k\to 0$,  
this resumed amplitude can be approximated by ${\cal A}_0^{\rm sum} \approx  - C_0(k )$,
therefore smoothly vanishes.
It is encouraging that infrared singularities disappear in this 
resumed amplitude.

The running coupling $C_0(\mu)$ depends on the boundary conditions 
$\Lambda$ and $C_0(\Lambda)$, each of which cannot be determined separately. 
Analogous to introducing $\Lambda_{\rm QCD}$ in Quantum Chromodynamics (QCD),  
it is useful to trade them for a new scale $\rho$:
\bqa
\rho & = &  \Lambda\, \exp\left[{4\pi\over m \, C_0(\Lambda)}\right]\, .
\label{rho:definition}
\eqa
One can easily verify that $\rho$ is RG invariant.

Because mass in the nonrelativistic theory is merely a passive parameter,
it can be transformed away, and consequently $C_0$ can be tuned dimensionless. 
In this sense, one observes that the nonrelativistic system with a $\delta^2({\bf r})$ potential
is classically scale-invariant~\cite{Bergman:1991hf}.
However, renormalization  necessarily generates a dynamic scale $\rho$,
thus breaks the scale invariance  quantum-mechanically.
This phenomenon is the manifestation of the so-called {\it dimensional transmutation}~\cite{Coleman:book}, 
also referred to as scale anomaly in Ref.~\cite{Bergman:1991hf}.

One can invert (\ref{rho:definition})  and express the effective coupling $C_0$ 
in term of $\rho$,
\bqa
C_0(\mu)&=&{4\pi\over m}\ln^{-1} \left({\rho\over \mu} \right) \,.
\label{true:C0:smallk}
\eqa
When $\mu$ approaches $\rho$,  the effective coupling $C_0$ becomes strong,
finally diverges at $\mu=\rho$.

Short-range interaction in two spatial dimensions can be classified into
two categories.
Firstly, for $C_0(\Lambda)>0$, which we refer to
as {\it repulsive} interaction,
the scale $\rho$ is larger (and can be much larger) than the cutoff $\Lambda$.
The coupling $C_0(\mu)$  monotonically decreases as $\mu$ descends from $\Lambda$.

Another case, $C_0(\Lambda)<0$,  referred to as  {\it attractive} interaction,
is more interesting\footnote{The word ``attractive" is
somewhat inaccurate, because at small $k$, the effective interaction in this case also
becomes repulsive.}. 
Here we have $\rho<\Lambda$.
When $\mu$ descends from $\Lambda$ to $\rho$, 
the coupling $C_0(\mu)$  drops to $-\infty$. 
This bears some resemblance with QCD,  where the strong coupling $\alpha_s(\mu)$
increases as $\mu$ decreases, finally  blows up near $\mu=\Lambda_{\rm QCD}$.
However, very different from QCD,  just across $\mu=\rho$  infinitesmally, 
the effective coupling abruptly jumps to $+\infty$ and then gradually diminishes
as $\mu$ further decreases.
Note that regardless of the sign of $C_0(\Lambda)$,
the effective interaction at sufficiently small momentum 
is always weakly repulsive.

It is well known that $\Lambda_{\rm QCD}$, as an integration constant,
cannot be  pinned down from QCD itself.
However, if QCD is indeed embedded in a more fundamental theory,
{\it e.g.} the Grand Unified Theory, then $\Lambda_{\rm QCD}$ can be 
unambiguously determined.

The same reasoning applies to our case too.
Since nonrelativistic theory is necessarily only an effective theory, 
$\rho$ can be determined once the more fundamental theory is known.  
For example, the $\phi^4$ theory specifies such a microscopic theory.
Perturbative expansion in this theory is governed by the factor $\lambda/8\pi m$.
For definiteness, let us take the coupling $\lambda=4\pi m$, 
which lies in the perturbative regime.
We then find $\rho \approx 2 e^8\, m  \approx 10^4\, m$,  which is several order-of-magnitude
larger than the scalar mass.
At a cursory glance, such a gigantic scale can hardly be associated with any 
reasonable nonrelativistic observables.

The resumed amplitude (\ref{3d:bubble}) can also be expressed
in term of $\rho$:
\bqa
{\cal A}_0^{\rm sum}  & = &  
- {4\pi\over m}  
\left[ \ln\left({\rho\over k}\right) + {i\pi\over 2} \right]^{-1} \,.
\label{reexpress:A:kappa}
\eqa
For the attractive case, it is possible to tune the momentum equal to $\rho$. 
At this specific momentum, the amplitude becomes purely absorptive
and exceedingly simple,  $8i/ m$.
This scale can be viewed as the transition point between the attractive
and repulsive interaction.

One can easily infer the imaginary pole $k=i\kappa$ of the amplitude.
It is nothing but $\kappa=\rho$.
Since $\rho$ is positive definite,  one may naively expect that
regardless of the sign of $C_0(\Lambda)$,  there always exists a bound state.
Nevertheless, recall for repulsive interaction,  $\rho > \Lambda$, 
which is (far) beyond the applicable range of the nonrelativistic effective theory.
Therefore, this bound state pole is fictitious and should not be endowed
with any physical significance.

On the contrary, for attractive interaction, 
$\rho$ is exponentially suppressed relative to $\Lambda$, therefore
there does exist  a  true bound state with binding energy 
\beq
E_B \approx -{\Lambda^2\over m}\,\exp \left[-{ 8\pi\over m |C_0(\Lambda)|}
\right].
\eeq
Therefore, the smaller $|C_0(\Lambda)|$ is, the much shallower the bound state becomes.
To be specific, let us again take the $\phi^4$ theory as an example.
If we allow $\lambda$ to be negative\footnote{Let us don't worry about the vacuum stability problem for a moment.},
and take $\lambda=-4\pi m$ for instance, 
we find the binding energy  $E_B \approx -4\, e^{-16}\, m \approx -10^{-7}\,m$,
corresponding to a rather shallow bound state.

One bonus comes from the RG equation.
The amplitude in (\ref{3d:bubble}), being RG invariant, enables us to 
organize the  logarithms of $\lambda^m\ln^n(2m/k)$
($m>n$) in the $\phi^4$ theory in a most efficient way.
After the tree-level matching (\ref{tree:C0}),  
we can  ascertain the leading logarithms $\lambda^{n+1}\ln^n k$ 
in $T_0$:
\bqa
 & & (-1)^{n+1} \lambda \left[ {\lambda \ln(2m/k) \over 16\pi m}  \right]^n\,.
\label{LL:phi4}
\eqa
They can also be directly inferred from the full theory,
which simply follow from the $s$-channel bubble chain.

Similarly, once the one-loop matching is done,
we are able to determine all the next-to-leading logs
of the form $\lambda^{n+2}\ln^n (2m/k)$.
Setting $\mu=2m$ in (\ref{3d:bubble}), 
plugging $C_0(2m)=\lambda/4m^2-(\lambda/4m^2)^2\,(m/2\pi)$ in, 
and expanding a first few terms, for example,
one can determine the coeffiecients of $\lambda^3 \ln k$ and $\lambda^4\ln^2 k$:
\bqa
 & &  -{\lambda^3 \ln(2m/k) \over (8\pi m)^2}\,
 \left( 1+{i\pi\over 4} \right)\,,
 \\
 & &  {\lambda^4 \ln^2(2m/k) \over 2 (8\pi m)^3}\,
 \left( 1+{3 i\pi\over 8} \right)\,.
 \label{NLL:phi4}
\eqa
These results can not be easily obtained  in the relativistic $\phi^4$ theory,
because one has to extract these logarithms from 
rather complicated two-loop and three-loop diagrams.

It is straightforward to generalize (\ref{3d:bubble}),
to include those higher-derivative opearators in the bubble chain sum.
Because the relation $I_n \approx I_0 k^{2n}$ holds regardless of 
the spacetime dimensions,
the same argument leading to the resummation formula (\ref{bubble:2d:all}) in 2D also applies here.
Therefore, the full bubble chain sum renders
\bqa
{\cal A}_0^{\rm sum}  & = &  
-\left[{1\over C_0 + C_2 \, k^2+\cdots}
\, + \, {m\over 4\pi} 
\left[ \ln\left({\mu\over k}\right) + {i\pi\over 2} \right]
\right]^{-1} \, .
\label{full:3D:nonrel:sum}
\eqa

We can infer the RG equation for $C_2$ from this equation through
a shortcut.
The first term in the bracket can be expanded,
$1/(C_0+C_2 k^2+\cdots)\approx 1/C_0-C_2/C_0^2\,k^2+O(k^4)$.
Notice that $1/C_0(\mu)$ together with the $\ln \mu$ term
forms a RG invariant.
The residual terms  must be $\mu$ independent
at any order of $k^2$ individually.
Consequently,  one can read off the RG equation of $C_2$:
\bqa
\mu{d\over d\mu} \left({C_2\over C_0^2}\right)  & =& \,0 .
\label{RG:C2}
\eqa
Therefore $C_2(\rho)$ diverges as $C_0^2(\rho)$.
One can verify that in general, $C_{2n}(\rho) \propto  C_0^{n+1}(\rho)$.

This RG equation can be confirmed  by directly working out the counterterms of $C_2$,
which arise from the bubble chain containing all $C_0$ vertices but one $C_2$ vertex:
\bqa
C_2^B &=& \mu^{2\epsilon} \left[C_2+ 2 \left({m\over 8\pi}\right) {C_2\,C_0\over \epsilon} +
3\left({m\over 8\pi}\right)^2 {C_2\,C_0^2\over \epsilon^2} +\cdots \right]
\nn \\
&=& {\mu^{2\epsilon}  \, C_2 \over \left(1-{m\over 8\pi} {C_0\over \epsilon}\right)^2 }\,.
\label{bare:C2:renormC2}
\eqa
Dividing this equation by the square of  (\ref{bareC0:renormC0}), one immediately
recovers (\ref{RG:C2}). 
With (\ref{beta:function:C0}) as the input, one readily deduces the $\beta$ function for $C_2$:
\bqa
\beta(C_2,\epsilon) 
&=& -2\epsilon \, C_2 + {m\over 2\pi}\, C_0\,C_2\,.
\label{beta:function:C2}
\eqa

Because the resumed amplitude (\ref{full:3D:nonrel:sum}) is RG invariant, 
we have the freedom to choose any $\mu$ which we prefer.
At first sight, setting $\mu=\rho$  leads to great simplification,
since all the $C_{2n}(\rho)$ diverge, so the first term in the bracket may be dropped.
Consequently, the amplitude still remains the very simple form of (\ref{reexpress:A:kappa}),
and the pole still remains at $\kappa=\rho$.

This is puzzling, for it implies that effects of those higher-derivative operators 
can be totally discarded.  A closer examination discloses that, 
choosing $\mu \approx \rho$ will instead lead to a highly unstable answer.
Recall $C_0$ diverges to either $+\infty$ or $-\infty$,  
depending on toward which direction $\mu$ approaches $\rho$. 
Similar pattern  occurs for $C_4$, $C_8$,  and so on.
Therefore,  for any finite $k$,  if one chooses $\mu$ very close to $\rho$,  
there is a possibility of large cancellations among different terms in the series $\sum C_{2n}\,k^{2n}$.  
As a result, $1/\sum C_{2n}(\rho)\,k^{2n}$ may not vanish as one naively expects.

A judicious analysis indicates that the higher-derivative operators do affect 
the location of the pole.
From  (\ref{full:3D:nonrel:sum}), one finds that the pole $\kappa$ 
no longer coincides with $\rho$, but shifts by an amount of 
\bqa
\Delta \kappa & = &  {4\pi \over m} \, {C_2(\mu)\over C_0^2(\mu)} \, \rho^3 + O(\rho^5)\,.
\label{norel:shift}
\eqa
Note this shift is RG invariant, as it must be.

The RG equation of $C_2$, (\ref{RG:C2}), indicates that $C_2(\mu)\sim \ln^{-2}(\rho/ \mu)$
as $\mu\to 0$,  vanishing in a more rapid speed than $C_0$. 
However, so far we have neglected the renormalization of $C_2$ by $C_0$ through
relativistic corrections. 
Recall in (\ref{3D:C2:oneloop:mat}), 
the relativistic effect tends to enhance  $C_2(\mu)$ as $\mu$ decreases,
which counteracts the effect represented by (\ref{RG:C2}).
The true RG flow of $C_2$ will depend upon the competition between them.

Now let us rederive the RG equation for $C_2$, this time
including  the effects of relativistic corrections.
The leading  relativity-induced counterterm 
can be extracted from $\tilde{I_0}$, and has been given in (\ref{counterterm:C2}). 
Higher-order counterterms can be worked out analogously by
computing the bubble diagrams which contribute at $O(k^2)$.
These diagrams can have $C_0$, $\delta C_0$  or lower-order $\delta C_2$
induced by the relativistic corrections,
as their vertices, and may also need one relativistic vertex insertion in the loops.
Adding these new counterterms to (\ref{bare:C2:renormC2}), we have
\bqa
C_2^B &=&  \mu^{2\epsilon} \left[ {C_2 \over \left(1-{m\over 8\pi} {C_0\over \epsilon}\right)^2 }
-{1\over 2(8\pi)m}\,{C_0^2\over \epsilon} 
-{1\over (8\pi)^2}\,{C_0^3\over \epsilon^2} 
-{3m\over 2(8\pi)^3}\,{C_0^4\over \epsilon^3} -\cdots
\right]
\nn \\
&=&  \mu^{2\epsilon} \: { C_2 - {1\over 2(8\pi)m}\,{C_0^2\over \epsilon} 
 \over \left(1-{m\over 8\pi} {C_0\over \epsilon}\right)^2 } \,.
\label{bare:C2:true:renormC2}
\eqa
Note these new counterterms can also be cast into a geometric series.
Acting $\mu\, d/d\mu$ on this equation, applying the chain rule,
we  obtain the full $\beta$ function for $C_2$:
\bqa
\beta(C_2,\epsilon) & =& -2\epsilon \, C_2+{m\over 2\pi}\,C_0 C_2 
- {C_0^2\over 8\pi m}\, .
\label{beta:C2:complete}
\eqa
In deriving this, the knowledge of $\beta(C_0,\epsilon)$ is needed.
Two competing  forces driving  the RG flow of $C_2$ are manifest in
this equation.
We pause to point out one subtlety concerning these two contributions.
Whereas the RG equation for $C_2$ obtained alone from the bubble chain 
consisting of all $C_0$ vertices except one $C_2$ vertex,
(\ref{beta:function:C2}), is self-consistent, 
the converse is not true.
It can be easily checked, if one keeps only the counterterms 
induced by the relativistic corrections in (\ref{bare:C2:true:renormC2}), 
no sensible $\beta$ function will be obtained\footnote{This means $\beta$ function
will contain the uncancelled poles.}.
Therefore, the relativistic effects cannot be isolated from
the higher-derivative operators.

Dividing both sides of this equation by $C_2$, 
one can arrange it into the form:
\bqa
\mu{d\over d\mu} \left({C_2\over C_0^2}\right) &=& - {1\over 8\pi m}\, ,
\label{RG:C2:rel:change}
\eqa
which can be easily solved:
\bqa
{C_2(\mu)\over C_0^2(\mu)} &=& {C_2(\rho)\over C_0^2(\rho)} +
{1\over 8\pi m}\ln\left({\rho\over \mu} \right).
\label{true:C2:behavior}
\eqa
We now see that, 
contrary to (\ref{RG:C2}), $C_2/C_0^2$ is no longer
a constant, but increases logarithmically as the renormalization
scale gets lower. 
In the $\mu \to 0$ limit, the RG flow of $C_2(\mu)$ is dominated 
by the relativistic corrections.
Keeping only the second term in the right-hand side, one finds
\bqa
C_2(\mu)  & \approx &  
 {2\pi\over m^3}\ln^{-1} \left({\rho\over \mu} \right) \, .
\label{true:C2:smallk}
\eqa
Therefore,  $C_2$ approaches  zero in the same speed as $C_0$.
Comparing (\ref{true:C0:smallk}) and (\ref{true:C2:smallk}), 
one finds a universal relation irregardless of any  specific planar system: 
$C_2(\mu)/C_0(\mu)\approx 1/2m^2$  at sufficiently small $\mu$.

The concise form of (\ref{bare:C2:true:renormC2}) suggests
that the relativistic effects can also be incorporated in
the resumed amplitude.
Similar to the consideration leading to (\ref{bubble:2d:rel}) in 2D, 
such a resummation formula must first reproduce the
one-loop amplitude (\ref{A0:3D:1loop}), which includes 
the first-order relativistic correction.
We  thus generalize (\ref{full:3D:nonrel:sum}) to
\bqa
{\cal A}_0^{\rm sum}  & = &  
-\left[{1\over C_0 + C_2 \, k^2+\cdots}
\, + \, {m\over 4\pi} \left( 1-{k^2\over 2 m^2} +\cdots \right)
\left[ \ln\left({\mu\over k}\right) + {i\pi\over 2} \right]
\right]^{-1} \, .
\label{full:3D:rel:sum} 
\eqa
Expanding the first term in the bracket to $O(k^2)$,  combining it with the
$k^2\ln \mu$ term, and demanding them to be  $\mu$  independent,
we  recover the RG equation for $C_2$, (\ref{RG:C2:rel:change}). 
This provides a cogent  support for this  formula.

We need  to know the higher-order relativistic corrections. 
The $\phi^4$ theory again provides the useful guidance in helping to recognize the pattern.
Examining (\ref{3d:s-channel:integr}), 
one finds that, interestingly enough,  
this series is again represented by the dilation factor $\gamma^{-1}$.
One can verify this resummation formula by all kinds of straightforward computations.

Knowing the sturcture of the exact amplitude,
we can pin down those logarithms accompanying $k^2$ in the relativistic $\phi^4$ theory,
analogous to what we have done in (\ref{LL:phi4})--(\ref{NLL:phi4}).
After the tree-level matching,
one can  infer the leading logarithms  $\lambda^{n+1} k^2 \ln^n k$ 
in $T_0$ to all orders:
\bqa
 & & (-1)^n \, {n \, k^2\over 2m^2} \, \lambda 
 \left[ {\lambda\,\ln(2m/k) \over 16\pi m}  \right]^n\,.
\eqa
In the full theory, these leading logarithms come from 
the $s$-channel bubble chain  with the first-order relativistic correction retained.

Once $C_0$ and $C_2$ are determined through the one-loop matching,
we are able to know all the next-to-leading logs
of the form $\lambda^{n+2} k^2  \ln^n k$.
Taking $\mu=2m$ in (\ref{full:3D:rel:sum}), 
substituting $C_0(2m)$ and $C_2(2m)= (\lambda/4m)^2/(48\pi m^3)$ in, 
and expanding a first few terms, for example, one can determine
the next-to-leading logs at  $O(\lambda^3)$ and $O(\lambda^4)$:
\bqa
 & &  {k^2\over m^2} \, {\lambda^3 \ln(2m/k) \over (16\pi m)^2}
 \left( {13\over 6} +  i\pi \right) \,,
 \\
 & &  - {k^2 \over 4 m^2} \, {\lambda^4 \ln^2(2m/k) \over (16\pi m)^3}\,
 ( 25 + 9 i\pi)\,.
\eqa
Needless to say, these results are difficult to derive in the full theory.

When relativistic effects are included,  the pole shifts from $\rho$ by an amount of
\bqa
\Delta \kappa & = &  {4\pi \over m} \, \left[{C_2(\mu)\over C_0^2(\mu)} -{1\over 8\pi m}
\ln\left( \rho \over \mu \right)
\right] \,  \rho^3 + O(\rho^5)
\nn \\
&=&  {4\pi \over m} \, {C_2(\rho)\over C_0^2(\rho)} \, \rho^3+ O(\rho^5)\,,
\eqa
where we  resort to (\ref{true:C2:behavior}) in the second line.
Evidently, this shift is also RG invariant, and much resembles its counterpart 
without incorporating  relativistic corrections, (\ref{norel:shift}).
The correspoinding binding energy then becomes
\bqa
E_B 
& = &
-{\rho^2 \over m} \left[ 1 +  {8\pi \over m}\, {C_2(\rho)\over C_0^2(\rho)} \, \rho^2
+ {\rho^2 \over 4 m^2} + O(\rho^4) \right] \,.
\eqa

Requiring the resumed amplitude (\ref{full:3D:rel:sum}) to be RG invariant, 
we can infer the RG equations for all remaining Wilson coefficients. 
Let us take $C_4$ as a specific example.
Expanding $1/\sum C_{2n}\,k^{2n}$  and  $\gamma^{-1}\ln\mu$  to the 4th order of $k$, 
piecing their $O(k^4)$ coefficients together and
demanding  it to be $\mu$ independent,
we obtain the following coupled RG equation:
\bqa
\mu{d\over d\mu} \left( {C_4\over C_0^2} - {C_2^2\over C_0^3} \right) 
&=&  {3\over 32\pi m^3}.
\label{RG:C4:full}
\eqa
Consequently, the $\beta$ function for $C_4$ can be readily identified:
\bqa
\beta(C_4) & =&  {m\over 2\pi}\,C_0 \,C_4 + {m\over 4\pi}\,C_2^2
- {C_0\, C_2\over 4\pi m}  + {3\, C_0^2\over 32\pi m^3}\,.
\label{beta:C4:complete}
\eqa
The first two terms in the right-hand side are as expected from (\ref{full:3D:nonrel:sum}),
when the relativistic effects are turned off.
The third term arises from implementing the first-order relativistic correction in bubble diagrams
containing all $C_0$ vertices but one $C_2$ vertex, whereas
the last term stems from 
the second-order relativistic correction in the bubble diagrams comprising 
entirely of $C_0$ vertices.

From (\ref{RG:C4:full}) and the asymptotic behaviour of $C_0$ and $C_2$, 
one  finds that $C_4(\mu)$  in the $\mu\to 0$ limit is approximately
\bqa
C_4(\mu)  & \approx &  
-  {\pi\over 2 m^5}\ln^{-1} \left({\rho\over \mu} \right).
\label{true:C4:smallk}
\eqa

We are now at a stage to understand the general pattern of the asymptotic behaviour of 
$C_{2n}$ coherently. 
If we take $\mu=k$ in (\ref{full:3D:rel:sum}), the terms explicitly depending  upon $\ln \mu$ vanish.
In the $k\to 0$ limit, the amplitude is well approximated by $- \sum C_{2n}(k)\, k^{2n}$.

Because the amplitude is RG invariant, we can freely swith to a different $\mu$.
Recall we have warned that taking $\mu=\rho$  doesn't produce a stable result. 
However, at sufficient small $k$, 
it is permissible to choose $\mu$ around $\rho$ in (\ref{full:3D:rel:sum}).
In the $k\to 0$ limit,  one can approximate
$1/\sum C_{2n}(\rho)\, k^{2n}$  by $1/C_0(\rho)$, which hence vanishes.
The resumed amplitude therefore reduces to
\bqa
{\cal A}_0^{\rm sum}  & \approx &  
- {4\pi\over m} \, \gamma\,
\ln^{-1} \left({\rho\over k} \right) 
\,.
\label{A0:3d:complete:concise}
\eqa

Recall $\gamma=1+ k^2/2 m^2- k^4/8 m^4+\cdots$,  so we immediately identify 
the asymptotic forms of $C_{2n}(k)$, and readily reproduce the earlier results 
in (\ref{true:C0:smallk}), (\ref{true:C2:smallk}) and (\ref{true:C4:smallk}).
This signals, regardless of the boundary conditions $C_{2n}(\Lambda)$,
all the Wilson coefficients at sufficiently small renormalization scale $\mu$
are effectively generated by the relativistic effects--  all of them 
are inversely proportinal to $\ln(\rho/\mu)$, with the coefficients fixed by 
the dilation factor.

\subsection{Four dimensions}
\label{4d:phi4:subsection}

Because of its close connection with the reality, short-range force in three spatial dimensions 
has been extensively discussed in the literature.
Our main new result is to fully incorporate the relativistic effects in the resumed amplitude.
We also determine the effective range in the $\phi^4$ theory, which roughly equals the Compton wavelength.

\subsubsection{Matching of 4D $\phi^4$ theory}

In the 4D $\phi^4$ theory,  
the coupling $\lambda$ is dimensionless and this theory becomes renormalizable.
In addition,  because the spatial dimension is now big enough, 
the fixed-order perturbation series no longer suffers the 
zero-momentum singularity as encountered in  2D and 3D.

It is necessary to specify the renormalization prescription.
It is standard to use  MS  when studying the high energy process,
where the running coupling and running mass are valuable notions.
However, for the nonrelativistic problem at hand, 
it is most convenient to choose the {\it on-shell renormalization} scheme,
in which the renormalized coupling $\lambda$ and the mass $m$  
are physical observables.
In this scheme,  the counterterm $\delta \lambda$ is chosen such that
the 2-body amplitude in the zero-momentum limit
remains fixed at $-\lambda$ to all orders.

The $T$-matrix to one loop order in this scheme is~\cite{Peskin:ev}
\bqa
 T &=& \:  -\lambda -{\lambda^2 \over 32\pi^2} 
\int_0^1 \, dx \left\{
\ln \left[ {1-x(1-x)s/m^2-i\epsilon \over 1- 4x(1-x) } \right] +
\ln  [ 1-x(1-x)t/m^2 ] \right.
\nonumber \\ 
& & \quad + 
\left.
(t\rightarrow u) 
\right\} \, .
\label{Peskin}
\eqa

The $s$-channel integral can be worked out and expanded:
\bqa
\int_0^1 \: dx \, \ln[1- x(1-x)\,s/m^2-i\epsilon ] &=&
-2 + {4 k \over \sqrt{s} } 
\left(
\tanh^{-1} \beta - {i\pi \over 2}  \right) 
\nn \\
&\approx&  -2 + {2 k^2\over m^2} - {i\pi k\over m} 
\left( 1-{k^2\over 2m^2}\right)\,.
\label{4D:s-channel}
\eqa

One can carry out the $t$-, $u$-channel integrals, similar to as in 2D and 3D:
\bqa
\int_0^1 \: dx \, \left\{ \ln[1- x(1-x)\,t/m^2 ] + (t\rightarrow u) \right\}
&\approx&
-{t+u\over 6m^2} = {2k^2\over 3m^2}.
\label{4D:tu-channel}
\eqa
When calculating the $t$- and $u$-channel diagrams,
the authors of Ref.~\cite{Consoli:1997ip} neglect mass of the scalar particle 
in the loop integral. 
After Fourier-transforming the amplitude to the coordinate space, 
they then find  that two particles  effectively
experience a $-1/r^3$  long-range potential.
We should stress, nevertheless, 
the  approximation $m=0$ inside the loop is not legitimate. 
Since the typical virtuality of the internal momenta
is $O(m^2)$,  the Uncertainty Principle implies that
these virtual particles cannot propagates much farther than 
the Compton wavelength. 
Thus the effects of these diagrams should be mimicked by 
the local operators, instead of
by an  instantaneous, nonlocal potential.

Piecing (\ref{4D:s-channel}) and (\ref{4D:tu-channel}) together, 
we obtain the S-wave amplitude:
\bqa
 T_0^{\rm 1-loop} &=&  -\lambda +{\lambda^2 \over 32\pi^2}
\left[
{i\pi  k \over m} \left( 1-{k^2 \over 2 m^2}\right)
- { 8 k^2 \over 3m^2} 
\right]\,. 
\label{One-loop:expans}
\eqa
Absence of  constant term at $O(\lambda^2)$ is specific
to the on-shell renormalization scheme.

Next we consider the one loop calculation in the EFT sector.
Fig.~\ref{fig:match}b) is already known in (\ref{true:4d:oneloop}),
and we need simply to include the first-order relativistic correction:
\beq
\label{4d:oneloop:full}
I_0 = { m\over 8\pi}\,\sqrt{-2mE-i\epsilon}
 \approx -{ i m k\over 8\pi}\,\left(1-{k^2\over 8m^2}\right).
\eeq

Fig.~\ref{fig:match}c) and d),  
the one loop diagram with one insertion of the relativistic vertex,
can be obtained by substituting $D=4$ in (\ref{D:oneloop:rel}):
\bqa
\tilde{I_0} &=& 
{im k \over 8\pi}\,\left({3 k^2 \over 8m^2}\right).
\label{4d:oneloop:rel}
\eqa

Merging  these together,
we obtain the S-wave amplitude  in the EFT sector:
\bqa
 {\cal A}_0^{\rm 1-loop} &=&  -C_0 + i\,C_0^2\,{m k\over 8\pi} \, 
 \left( 1-{k^2\over 2m^2} \right) - C_2\,k^2.
\label{A0:4d:1loop}
\eqa

By construction,  $C_0$ doesn't receive any modification 
with respect to (\ref{tree:C0}).
Needless to repeat, both the full theory and the EFT share the same
non-analytic (imaginary) terms. 
The relativistic corrections only influence the imaginary parts of the amplitude.
Matching (\ref{One-loop:expans}) onto (\ref{A0:4d:1loop})
via (\ref{matching:formula}), we then read off $C_2$:
\bqa
 C_2 & = &  {1\over 3 m^4}\, \left({\lambda \over 4 \pi}\right)^2
 + O\left(\lambda^3\right).
\eqa
%

\subsubsection{Effective range expansion}

If the scattering length in a nonrelatitivistic system is of natural size,
we can simply stick to   MS in the EFT sector.
Contrary to the lower dimensional cases, 
it is not compulsive here to sum the bubble diagrams to all orders.
Retaining first few terms in the perturbation series suffice for practical purpose.
Nevertheless, it is still instructive to perform the full bubble sum.
We can routinely generalize (\ref{finite:4d:resum}) to incorporate all the 
higher-derivative terms, by replacing $1/C_0$ with 
$1/\sum C_{2n}\, k^{2n}$.

Unlike in 2D and 3D, there is no strong motivation to include  relativistic corrections. 
Nevertheless,  for completeness and clarification,
we proceed to give s resummation formula  which fully implements the relativistic effects.

Analogous to the previous analyses,   inspecting  the imaginary part of the one-loop 
$s$-channel integral in the $\phi^4$ theory, (\ref{4D:s-channel}), 
we find that the relativistic factor is again represented by $\gamma^{-1}$, 
exactly the same as in 2D and 3D. 
Therefore, the Lorentz-invariant resumed amplitude is
\bqa
 {\cal A}_0^{\rm sum} &=& 
 - \left[ {1\over C_0 + C_2 k^2 + \cdots  } + {im\over 8\pi}\,\gamma^{-1} k \right]^{-1} \,.
\label{full:geom:sum:4D}
\eqa
At small $k$, this amplitude is dominated by $-C_0$.  
Evidently, it is not essential to include the relativistic corrections.

A by-product of this resumed amplitude is that it conveniently 
embodies the optical theorem. 
One can quickly read off the uncalculated higher order immaginary part 
from the lower order results. 
For example, at $O(\lambda^3)$, the leading imaginary term in $T_0$ is
\beq
 2 i\, C_0 \,C_2\, {m \, k^3\over 8\pi}\, (4m^2) = {i\lambda^3\over 3 (4\pi)^3} \,
{k^3\over m^3}\, .
\eeq
It will be little bit laborious to infer from the relativistic $\phi^4$ theory.

According to the partial wave expansion, the S-wave partial amplitude can be 
parameterized as\footnote{The  kinematic factor is chosen such 
to reproduce the cross section formula  in the relativistic field theory:
${d\sigma\over d\Omega} = {1\over 64\pi^2 s} |T_0|^2$.  
The phase shift is defined according to  $ {d\sigma\over d\Omega}={4 \sin^2 \delta_0 \over k^2}$,
where the effect due to identical bosons is accounted by the factor 4.}
\bqa
{\cal A}_{0} &=&  {8\pi \over m}\: {\gamma\over k} \:
e^{i\delta_0}\, \sin \delta_0 \,,
\label{partial:wave:exp}
\eqa
where $\delta_0$ is the S-wave phase shift.
It is convenient to rewrite  $\exp(i\delta_0)\sin \delta_0 = 1/(\cot \delta_0 - i)$.
It is well known that in the low energy scattering, each partial amplitude 
is insensitive to the fine structure of the short-range  potential, 
instead can be characterized rather accurately by a few parameters only. 
This idea, akin to the spirit of EFT, is referred to as {\it effective range expansion}.
According to this ansatz, we parameterize the S-wave phase shift as\footnote{
Note the relativistic factor $\gamma^{-1}$ is absent in the standard definition.} 
\beq
 \gamma^{-1} \, k  \cot\delta_0 =
-{1 \over a_0} + {r_0\over 2} \, k^2 +\cdots\,,
\label{ERE}
\eeq
where $a_0$ is the S-wave scattering length, and $r_0$ is the effective range.
The S-wave amplitude can thereby be written  
\bqa
{\cal A}_{0}
&=& {8\pi \over m} \,
\left[ -{1 \over a_0} + {r_0 \over 2} \, k^2 +\cdots - i \gamma^{-1} k \right]^{-1} .
\label{Eff-Range:expans}
\eqa

Comparing (\ref{full:geom:sum:4D}) and (\ref{Eff-Range:expans}), one sees that
EFT and effective range expansion are completely equivalent.
Note they share the same structure of relativistic corrections.
Recalling the values of $C_0$ and $C_2$  from the one-loop matching,  
we can identify the scattering length and the effective range  in the $\phi^4$ theory:
\bqa
 a_0 &=& {m\over 8\pi}\, C_0 =  {\lambda \over 32\pi m}\,,
\nn \\
 r_0 & = &  {16\pi\over m}\, {C_2\over C_0^2}
= {16 \over 3\pi m}\,[1+ O(\lambda)]\,.
\label{boundary:condition:C0C2}
\eqa
Clearly, the on-shell renormalization scheme in the 
$\phi^4$ theory perfectly matches with the effective range expansion.
From this scheme, we acquire an exact scattering length,
and an approximate effective range, which  yet can be  expanded order by order in $\lambda$.
For the coupling lying in the perturbative region ($\lambda<16\pi^2$), 
we always have $a_0$  smaller than $r_0$.

It is interesting to note, the effective range in the $\phi^4$ theory at leading order
doesn't depend on $\lambda$.
It approximately equals the Compton wavelength,
consistent with what we have expected in Section~\ref{trivial}.
Because $k\,r_0\sim k/m \ll 1$, so practically it is unnecessary 
to include any higher partial waves.

One subtlety deserves being pointed out. If the factor $\gamma^{-1}$ were not absorbed
in the definition of the effective range expansion in  (\ref{ERE}),
$r_0$ would receive an additional correction,  $-32\pi/(\lambda m)$.
 This is an unacceptable situation,  
since a very weak coupling would correspond to a very large negative effective range!

We  have adopted (\ref{new:matching:Lagr}) to 
implement  relativistic corrections  in this work.
It is worth commenting on  what if the alternative scheme, 
(\ref{field:red:lag}), is instead used.
It should be of some interest,  since this scheme seems to be favored by many authors.
To convert the resumed formula into this scheme,  
one needs  divide (\ref{full:geom:sum:4D}) and (\ref{partial:wave:exp}) by $\gamma^2$, 
in compliance with the reduction formula (One can understand this factor by
comparing (\ref{matching:formula}) and (\ref{different:matching})). 
The corresponding amplitude in this scheme then reads:
\bqa
 {\cal A}_0^{\prime\,{\rm sum} } &=& 
 - \left[ {\gamma^2\over C_0 + C_2 \, k^2 + \cdots  } + {im\over 8\pi}\,\gamma k \right]^{-1} \,.
\label{full:geom:sum:4D:another:rel}
\eqa
Keeping the lowest order terms, one  readily recovers  ${\cal A}_0^{\prime \, {\rm tree}}$ 
in (\ref{tree:ampl:2nd:rel}).
We notice that Ref.~\cite{vanKolck:1998bw} presents a similar resummation formula
which contains the correct imaginary term in the bracket,  
but misses the factor $\gamma^2$  in the first term.  
Note those two-body operators induced by the field redefinition (\ref{field:redefine})
have not been included in Ref.~\cite{vanKolck:1998bw}.

Although triviality of the $\phi^4$ theory cannot be substantiated in the
nonrelativistic limit, it is  possible to establish a much weaker assertion-- that 
the strong interacting $\phi^4$ theory may not exist.

It  is well known that an unusually large positive (negative) $a_0$ corresponds to
a threshold (virtual) bound state.  This can be understood from (\ref{Eff-Range:expans}),
the pole $\kappa$ is located roughly at $1/a_0$ for $a_0\gg r_0$.  
For the $\phi^4$ theory to be well defined, we must request a positive $\lambda$, 
and the repulsive interaction forbids it to host any bound state.
It is equivalent to say that this theory cannot possess a large positive scattering length.
As a result,  the self-coupling $\lambda$ cannot be too strong.

Complimentary evidence comes from (\ref{boundary:condition:C0C2}).
If $\lambda$ can be very large, $r_0$ may significantly depart 
from the Compton wavelength, driven by the higher-order corrections.
This seems to conflict with the natural expectation that the effective range 
in this theory should always be of order $1/m$.

Let us quantify this discussion little bit.  The location of bound state pole 
for general $a_0$, $r_0$ can be  solved from (\ref{Eff-Range:expans}):
\bqa
 \kappa &=& {1\over a_0}\, {2\over 1+ \sqrt{1-2r_0/a_0}}\, ,
\eqa
where we neglect the insignificant relativistic correction.
To nullify such a pole,
one requires $a_0 < 2 r_0$,  so that $\kappa$  doesn't  admit a real solution.
This imposes a bound $\lambda< 32^2/3\approx 341$.
Unfortunately, this bound is too loose to be useful,  because when $\lambda$
exceeds $16\pi^2\approx 158$, the perturbative matching  can  no  longer be trusted, 
neither can the effective range determined from it.

\subsubsection{Incorporating Relativity in PDS}

The PDS scheme is tailor-made for describing the 
finely-tuned system in which the scattering length becomes unnaturally large~\cite{Kaplan:1998tg}.
In this scheme, 
one not only subtracts the $D=4$ pole as in MS,
but also the $1/(D-3)$ pole which corresponds to the linear divergence at $D=4$.
As a result,  the subtracted integral  depends  on the subtraction scale $\mu$ linearly, 
which mimics the effects of $\Lambda$ in the cutoff scheme.

The integral $I_n$ in (\ref{good:loop:feature}) exhibits a  $D=3$ pole,
which can be removed by adding to $I_n$ the counterterm
\bqa
\delta I_n &=& - \left({m\over 8\pi} \right) {(2Em)^n \, \mu    \over D-3} \, ,
\eqa
so that the subtracted integral in $D=4$  is
\bqa
I_n^{\rm PDS} = I_n+ \delta I_n &\approx& - \left({m\over 8\pi}\right) \, k^{2n}\, (\mu + ik)
\, ,
\eqa
where the relativistic correction has been neglected.

Since the relation $ I_n\approx I_0\, k^{2n}$ still holds in  PDS,
according  to the preceding discussions,
the bubble chain diagrams incorporating all the higher-derivative terms 
can  still be summed analytically, and the result is
\bqa
 {\cal A}_0^{\rm sum} &=& 
 - \left[ {1\over C_0 + C_2\, k^2 + \cdots  } + {m\over 8\pi}\,(\mu+ ik) \right]^{-1} \,.
\label{geom:sum:4D:PDS:norel}
\eqa
At $\mu=0$,  PDS coincides with MS.

Requiring the amplitude to be $\mu$ independent, one can infer the RG equations for 
the Wilson coefficients.
For instance,  $C_0$ and $C_2$ satisfy the RG equations:
\bqa
{d  \, C_0\over d\mu}   & = & {m\over 8\pi} \, C_0^2  \, ,
\\
{d\over d\mu} \left({ C_2\over C_0^2}\right)  & = & 0 \,.
\eqa
Note they are identical to their counterparts in 3D,
(\ref{beta:function:C0}) and (\ref{RG:C2}), except one needs replace
$\mu$ by $\ln \mu$ and double the right side of both equations.
This is not unexpected,  because the same $1/(D-3)$ poles get subtracted
in both cases.

The solutions of these RG equations are
\bqa
C_0(\mu)  & = &  {8\pi \over m} \left({1\over a_0} -\mu \right)^{-1} \,,
\label{PDS:C0}
\\
C_2(\mu)  & = &  {4\pi r_0 \over m} \left({1\over a_0} -\mu \right)^{-2} \,.
\label{PDS:C2}
\eqa
The scattering length and effective range enter as the boundary condition,
which specify the effective coupling at $\mu=0$ through (\ref{boundary:condition:C0C2}).
Note $1/a_0$  here plays the similar role as $\rho$ in 3D. 
First,  the bound state pole is  approximately  located at $\kappa \approx 1/a_0$;
second,  all the couplings $C_{2n}$ diverge at $\mu=1/a_0$.
Nevertheless, contrary to the logarthmic running in 3D,  
the effective couplings in PDS depend linearly on $\mu$, so run quite fast.
When the momentum is larger than $1/a_0$, one usually chooses $\mu \sim k$, 
so that all the Wilson coefficients  have definite scaling in momentum, 
$C_{2n}(\mu)\sim 1/\mu^{n+1}\sim 1/k^{n+1}$~\cite{Kaplan:1998tg}.

Let us now take relativistic effects into account.
First consider the one loop integral with one relativistic vertex insertion, $\tilde{I_0}$,
whose D-dimensional expression is given in (\ref{D:oneloop:rel}). 
To get rid of its $D=3$ pole, we add the counterterm
\bqa
\delta \tilde{I_0} &=& \left({m\over 8\pi}\right) \, {k^2 \over 2m^2 } \, {\mu \over D-3} \,.
\eqa

Therefore, the one-loop integral $I_0+\tilde{I_0}$  is
\bqa
(I_0 + \tilde{I_0})^{\rm PDS}
&=& - {m\over 8\pi} \left(1- {k^2\over 2m^2} \right) \, (\mu + ik ) \,.
\eqa
It is identical to its MS counterpart,  (\ref{A0:4d:1loop}),
except $ik$ there should be promoted  to $\mu+ik$.

One can check that this pattern is completely general.  
Therefore,  the resummation formula (\ref{geom:sum:4D:PDS:norel}) can be generalized into
a Lorentz-invariant form:
\bqa
 {\cal A}_0^{\rm sum} &=& 
 - \left[ {1\over C_0 + C_2 \, k^2 + \cdots  } + {m\over 8\pi}\,\gamma^{-1} (\mu+ ik) \right]^{-1} \,.
\label{full:geom:sum:4D:PDS:rel}
\eqa
This accomplishment  should not bring much surprise. 
At any rate, PDS,  like MS, is based on the DR, 
and it is  well known that DR  preserves the spacetime symmetries by default.

The RG equation of $C_0$  is not affected by relativistic corrections.
After incorporating the relativistic effects,
$C_2$  satisfies the following RG equation:
\bqa
{d\over d\mu} \left({ C_2\over C_0^2}\right)  & = & - {1\over 16\pi m} \, ,
\eqa
which is again very similar to its 3D counterpart, (\ref{RG:C2:rel:change}).
It can be easily solved:
\bqa
C_2(\mu)  & = &  {4\pi \over m} \left(r_0- {\mu\over m^2} \right) 
\left({1\over a_0} -\mu \right)^{-2} \,.
\eqa
For any reasonable subtraction scale $\mu$, 
the relativistic correction, which  is represented by  $\mu/m$,
is always much smaller that  $r_0$.

Let us take the two-nucleon S-wave scattering as an concrete example,
to estimate the importance of the relativistic effects.
The effective range $r_0$ is roughly about $1/m_\pi$.
In fitting low-energy scattering data, one usually chooses $\mu=m_\pi$~\cite{Kaplan:1998tg}. 
One finds the relativistic effect reduces $C_2(m_\pi)$ given in (\ref{PDS:C2})
by  $m^2_\pi/m_N^2\approx 2\%
$.  As expected, this is a quite small effect.

\section{Summary}
\label{summary}

In this work,  we have presented a rather detailed study of
the short-range interaction in the 2-body sector. 
This study is expedited by employing the EFT approach,
which is in many aspects superior to the quantum mechanical formalism.

Considerable effort is devoted to clarifying some confusion concerning the
short-range interaction.
Based on the Uncertainty Principle, we argue that,  
any distance scale which can be confidently referred 
in nonrelativistic quantum mechanics must exceed the Compton wavelength.
The effective range should obey this criterion.
Therefore,  
contact interaction, such as the $\delta^3({\bf r})$ potential,  should be viewed
as an idealized, but  unrealistic and physically-irrelevant  notion.
The tenet can  be also stated in another way-- 
the cutoff $\Lambda$ in all the EFTs should be kept finite.
Evidently, triviality of the 4D $\phi^4$ theory cannot be
substantiated in the nonrelativistic limit.  
Nevertheless, from other considerations elaborated at the end of Sec.~\ref{matching:oneloop},
we argue that a much weaker assertion may hold true--  there is no
strongly interacting 4D $\phi^4$ theory.

There are very few exactly soluble models in quantum field theory.
Notably, the expression we obtained for the S-wave amplitude doesn't
involve any approximation.
This knowledge allows us to unequivocally extract important nonperturbative information
such as  the bound state pole, 
which will never show up  at fixed order perturbation series.
Although  the short-range force is basically a quantum mechanical problem, 
it is crucial to employ the field-theoretic method to accomplish this.
Therefore, this complete solution 
adds some wealth to the treasury of quantum field theory.

We have considered the short-range interactions in various spacetime dimensions.
Among them,  the attractive interaction in 3D is especially interesting. 
It shares some similar features with QCD,  {\it e.g.}, most notably, dimensional transmutation.
In this problem, a dynamical scale $\rho$,  which could be much smaller compared to 
the cutoff scale, is generated.
Analogous to the low-energy dynamics of QCD, which is mainly governed by $\Lambda_{\rm QCD}$, 
the dynamics in this case is largely controlled by $\rho$. 
In addition, 
this theory also displays asymptotic freedom in the region $\rho <\mu < \Lambda$.

Relativistic effects are usually thought unimportant in the nonrelativistic limit.
However, as we have seen in 3D,  relativity plays an important role in governing
the RG flows of the higher-dimensional operators at infrared scale.

The EFT method sheds useful light on the behaviour of 
the $\phi^4$ theory in the nonrelativistic limit.
At any fixed-order perturbative expansion, 
the 2D and 3D $\phi^4$ theories are  plagued by infrared singularities.
It is only with recourse to the EFT
that one can achieve a sensible result at small momentum.
Furthermore, the power of EFT is vividly exemplified by the renormalization group.
With the aid of the RG equations in 3D EFT,  
one can easily deduce the next-to-leading logarithms in the 3D $\phi^4$ theory. 
It is much more efficient than directly extracting them from multi-loop diagrams in the full theory.

The equivalence between the effective range expansion and 
the resummation formula in 4D EFT 
is  carefully  verified,  with the relativistic effects fully accounted for.
We then pinpoint the effective range in the 4D $\phi^4$ theory to be approximately $16/3\pi m$.
This nonzero result  provides the compelling evidence to our earlier statement, 
that no any physical system accommodates a zero-range interaction.

It is interesting, but challenging to infer the $O(\lambda)$ correction to the effective range.
To accomplish this,  the two-loop matching, hence the two-loop calculations in the $\phi^4$ theory 
are requested.
The two-loop integrals in the full theory are enormously complicated, unlikely to 
be worked out in a closed form.
Fortunately,  knowing their approximate expressions, which are expressed in power expansion of $k$,  
will be sufficient for the purpose  of matching.
The threshold expansion method~\cite{Beneke:1997zp} can be called for to fulfill this goal.

We have only considered  the two-body scattering in this work.
An interesting application of the EFT method is to study the many-body phenomena. 
For instance, the ground-state energy density for a dilute homogeneous gas 
have been calculated in the EFT framework~\cite{Braaten:1996rq,Hammer:2000xg}.
One can exploit similar techniques to attack the collective phenomena 
in one and two spatial dimensions.

\section*{Acknowledgments}

I am grateful to J.-W. Chen, H.~W. Hammer and U.~van Kolck for useful discussions 
at the early stage of this work. I thank E.~Braaten for the comment on the manuscript.
This work is supported in part by the 
National Science Foundation under Grant No. PHY-0100677.

\end{document}